*Review Article*

# Video Capsule Endoscopy and Ingestible Electronics: Emerging Trends in Sensors, Circuits, Materials, Telemetry, Optics, and Rapid Reading Software

**Dylan Miley,[1] Leonardo Bertoncello Machado,[1] Calvin Condo [iD],[1] Albert E. Jergens [iD],[2] Kyoung-Jin Yoon [iD],[3] and Santosh Pandey [iD][1]**

[1]*Department of Electrical and Computer Engineering, Iowa State University, Ames, Iowa, USA*
[2]*Department of Veterinary Clinical Sciences, College of Veterinary Medicine, Iowa State University, Ames, Iowa, USA*
[3]*Veterinary Diagnostic and Production Animal Medicine, College of Veterinary Medicine, Iowa State University, Ames, Iowa, USA*

Correspondence should be addressed to Santosh Pandey; pandey@iastate.edu

Real-time monitoring of the gastrointestinal tract in a safe and comfortable manner is valuable for the diagnosis and therapy of many diseases. Within this realm, our review captures the trends in ingestible capsule systems with a focus on hardware and software technologies used for capsule endoscopy and remote patient monitoring. We introduce the structure and functions of the gastrointestinal tract, and the FDA guidelines for ingestible wireless telemetric medical devices. We survey the advanced features incorporated in ingestible capsule systems, such as microrobotics, closed-loop feedback, physiological sensing, nerve stimulation, sampling and delivery, panoramic imaging with adaptive frame rates, and rapid reading software. Examples of experimental and commercialized capsule systems are presented with descriptions of their sensors, devices, and circuits for gastrointestinal health monitoring. We also show the recent research in biocompatible materials and batteries, edible electronics, and alternative energy sources for ingestible capsule systems. The results from clinical studies are discussed for the assessment of key performance indicators related to the safety and effectiveness of ingestible capsule procedures. Lastly, the present challenges and outlook are summarized with respect to the risks to health, clinical testing and approval process, and technology adoption by patients and clinicians.

## 1. Introduction

The gastrointestinal (GI) tract is both an intriguing and elusive environment for diagnostic procedures and therapeutic interventions. Its broad repository of electrolytes, metabolites, enzymes, and microbes serves as potential targets for several diseases and disorders [1, 2]. In this regard, real-time monitoring of gastrointestinal biomarkers in a safe, unobtrusive, and cost-effective manner seems achievable in the near future—in part because of the concerted growth, commercialization, and adoption of video capsule endoscopy and ingestible electronics [2–4].

A recent survey indicates that the ingestible capsule market is expected to be worth $8.98 billion by 2024 with key application areas in capsule endoscopy, remote patient monitoring, and targeted drug delivery [5]. Other drivers for the ingestible capsule market are the patient preference for minimally invasive procedures, favorable medical reimbursement policies, and the rising rate of stomach and colon cancers [5]. Modern ingestible capsules can navigate the gastrointestinal tract and inspect the spatial and temporal landmarks pertinent to common gastrointestinal disorders. To enable the diagnostic and therapeutic functions, a number of advanced features are embedded in modern ingestible capsules, such as microrobotics, targeted stimulation and neuromodulation, controlled sampling and delivery, fine needle biopsy, closed-loop control, panoramic imaging, and rapid reading algorithms [6–9]. The assessment of capsule functions is done using key performance indicators (KPIs) of accuracy and resolution, repeatability and robustness, autonomous navigation, power consumption, form factor, and operational ease. Besides these KPIs, a central

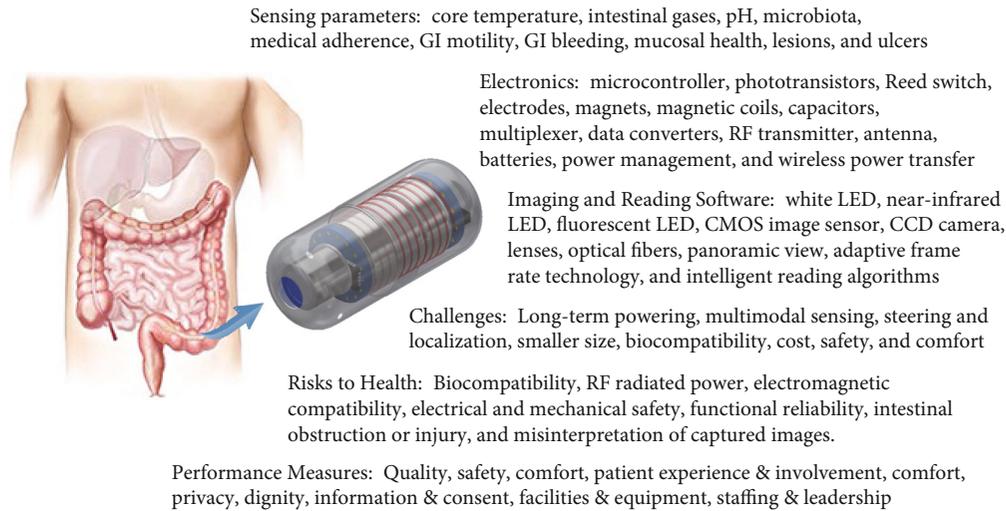

Sensing parameters: core temperature, intestinal gases, pH, microbiota, medical adherence, GI motility, GI bleeding, mucosal health, lesions, and ulcers

Electronics: microcontroller, phototransistors, Reed switch, electrodes, magnets, magnetic coils, capacitors, multiplexer, data converters, RF transmitter, antenna, batteries, power management, and wireless power transfer

Imaging and Reading Software: white LED, near-infrared LED, fluorescent LED, CMOS image sensor, CCD camera, lenses, optical fibers, panoramic view, adaptive frame rate technology, and intelligent reading algorithms

Challenges: Long-term powering, multimodal sensing, steering and localization, smaller size, biocompatibility, cost, safety, and comfort

Risks to Health: Biocompatibility, RF radiated power, electromagnetic compatibility, electrical and mechanical safety, functional reliability, intestinal obstruction or injury, and misinterpretation of captured images.

Performance Measures: Quality, safety, comfort, patient experience & involvement, comfort, privacy, dignity, information & consent, facilities & equipment, staffing & leadership objective of capsule technologies lies in improving the safety and quality of procedures while prioritizing patient comfort and involvement [10]. However, capsule system manufacturers face multiple hurdles as they move up the technology readiness levels (TRLs). Some challenges arise from the fact that human clinical trials are demanding, time-consuming, and expensive and often compounded by the issues of interpatient variability, disease complexities, and nonoptimal procedural parameters [11]. As such, our work is motivated by the competing tradeoffs within the field of ingestible capsules—the exuberance of research breakthroughs balanced by the rigors of clinical testing, approval process, and business sustainability [12–14].

Figure 1: Scope of the review. This work surveys the hardware and software technologies for video capsule endoscopy and ingestible electronics. We discuss the sensing parameters, electronics, imaging and reading software, challenges, risks to health, and performance measures. The image of the human gastrointestinal tract is reproduced (adapted) from Selber-Hnatiw S. et al. Human gut microbiota: toward an ecology of disease, Frontiers in Microbiology, 8:1265 (2017) under the Creative Commons Attribution License (CC BY).

Here, our purpose is to provide a comprehensive review of video capsule endoscopy and ingestible electronics with a focus on recent developments in hardware and software technologies (Figure 1). Section 1 introduces the structure, functioning, and disorders of the mammalian gastrointestinal tract. The time evolution of ingestible capsules and the guidelines for their safety and effectiveness are covered. Section 2 explains the capsule components for gastrointestinal sensing, imaging, and wireless telemetry with examples of ingestible sensors, devices, and circuits. Section 3 demonstrates the techniques for capsule navigation and localization in the gastrointestinal tract using magnetic actuation, robotic manipulators, and electrical stimulation, in addition to capsule placement devices. Section 4 illustrates the capsules to monitor the physiological status, including core body temperature, pH and gastroesophageal reflux, and intestinal gases. Section 5 shows the capsules to induce mechanical vibrations and bioelectric neuromodulation in the gut. Section 6 deliberates the capsules to track medication adherence, especially during self-administration of drugs at home. Section 7 presents experimental capsules for gut sample collection, bacterial biomarker detection, and microbiome analysis. Section 8 gives examples of video capsule endoscopes and rapid reading software, X-ray imaging and microultrasound capsules, and imaging-guided therapeutics and gastrointestinal surgical procedures. Section 9 highlights the trends in biocompatible materials, edible electronics, batteries, and alternative energy sources for the next-generation ingestible capsules. Section 10 discusses the present challenges and outlook for this field.

1.1. Mammalian Gastrointestinal Tract. The structure and functions of the mammalian gastrointestinal tract are dependent on the body and size of the organism, nature and frequency of food intake, need for food storage, and adaptation needs [15]. The human gastrointestinal tract is approximately 8 m long and comprises the oral cavity, esophagus, stomach, small intestine (duodenum, jejunum, and ileum), and large intestine (cecum, colon, and rectum). Table 1 lists the anatomic units of the human gastrointestinal tract with its respective dimensions, pH, maximum stress, disorders, and diagnostic procedures [8, 16–19]. It is noteworthy that the anatomy of the human gastrointestinal tract varies from person to person, and the physical dimensions of the anatomic units vary significantly with age, sex, gender, ethnicity, etc. The esophagus (pH = 5-6) has an approximate length of 18-25cm, diameter of 2.5 cm, and is accessible by endoscopy [16]. The stomach has an approximate length of 20 cm, diameter of 15 cm, and volume capacity of 1-1.6 liters. The stomach is very acidic (pH = 1.2-3.5) where the supply of hydrogen ions is provided by the secretions from parietal cells. The small intestine has an approximate length of 6.25 meters, diameter of 2.5 cm, and a curvy and convoluted shape. The acidic content of the stomach is neutralized in the duodenum (pH = 4.6-6.0) by alkaline fluid secretions (rich in bicarbonate ions) from the pancreas stimulated by secretin and primary bile acids produced from cholesterol in the liver. The small intestine is adapted for nutrient absorption by having a large surface area with villi which project into the luminal cavity. The effective surface area per unit villus surface is roughly around $25\,\mu m^2$ [15]. The large intestine is approximately 1.5 m long with a diameter of 6.5 cm and is involved in the recovery of water and electrolytes, formation and storage of feces, and fermentation of indigestible food matter by bacteria. The ileocaecal valve controls the passage of feces from the distal small intestine (ileum) into the colon. The colon is home to the largest number of bacterial microbes that produce a number of beneficial metabolites including short chain fatty acids to maintain intestinal health [20, 21].

Table 1: Structure and functions of the human gastrointestinal tract.

| Structure | Subsections | Size ($D \times L$), wall thickness | pH, maximum stress (MPa) | GI disorders and diseases | Diagnostic procedures |
|---|---|---|---|---|---|
| Esophagus | N/A | 2.5 cm × 18-25 cm, 4.7 ± 0.25 mm | pH: 5-6, 2.19 ± 0.06 (longitudinal), 1.41 ± 0.05 (circumferential) | Achalasia, Barrett's esophagus, esophageal cancer, GERD, peptic stricture, webs and rings | Esophagram, X-ray, gastroscopy, tissue biopsy, esophageal manometry, pH probe |
| Stomach | N/A | 15 cm × 20 cm, 3.92 ± 0.16 mm | pH: 1.2-3.5, 0.67 ± 0.19 (longitudinal), 0.5 ± 0.12 (circumferential) | Gastritis, gastroenteritis, gastroparesis, nonulcer dyspepsia, peptic ulcers, gastric cancer | pH monitoring, X-ray, gastroscopy, endoscopic ultrasound, gastric monometry, *Helicobacter pylori* breath test |
| Small intestine | Duodenum<br>Jejunum<br>Ileum | 2.5 cm × 26 cm, 1.5 ± 0.6 mm<br>2.5 cm × 250 cm, 1.5 ± 0.5 mm<br>2.5 cm × 350 cm, 1.5 ± 0.5 mm | pH: 4.6-7.6, 0.55 ± 0.33 (longitudinal), 0.92 ± 0.48 (circumferential) | Amyloidosis, celiac disease, dysmotility, inflammatory bowel disease, infections, intestinal lymphoma, lactose intolerance, tumors, small bowel obstruction | Small bowel manometry, X-ray, endoscopic ultrasound, endoscopic retrograde cholangiopancreatography (ERCP), hydrogen breath test, lactose tolerance test |
| Large intestine | Cecum<br>Ascendent colon<br>Transverse colon<br>Descending colon<br>Colon sigmoid | 9 cm × 6 cm, 4.8 ± 0.8 mm<br>6.6 cm × 20.9 cm, 5.0 ± 2.0 mm<br>5.8 cm × 50.2 cm, 4.2 ± 0.7 mm<br>6.3 cm × 21.8 cm, 3.2 ± 0.6 mm<br>5.7 cm × 38.3 cm, 3.2 ± 0.7 mm | pH: 7.5-8.0, 1.19 ± 0.30 (longitudinal), 065 ± 0.16 (circumferential) | Anal stenosis, colon cancer, colonic intertia, Crohn's disease, diverticulosis, inflammatory bowel disease, irritable bowel syndrome, rectal descent and prolapse, polyps, polyposis | Barium enema, x-ray, sigmoidoscopy, colonoscopy, capsule endoscopy, anorectal manometry, stool acidity test, fecal occult blood test, fecal immunochemical test |

Compared with the human alimentary tract, the gastrointestinal tracts of common laboratory animals have certain similarities and differences in their physiological, biochemical, and metabolic signatures [15, 22]. For instance, cyclic gastric motility is exhibited in both humans and common laboratory animals where stomach contractions aid in the mixing and grinding of stomach contents for entry into the duodenum. Humans and dogs have similar gastric morphologic features (i.e., glandular structure lined with cardiac, gastric, and pyloric mucosa), emptying characteristics, and a similar gut microbiome [23]. The dog is also an excellent model for investigation of gastrointestinal diseases [24]. Despite many anatomical difference of alimentary system between humans and pigs [25], pigs are considered a better model for translational research of human gastrointestinal tract regarding function, nutrition, and digestive diseases than many other laboratory models, particularly rodents [26–28]. Besides the fact that pigs are an omnivore like humans, the microbiota profile of the pig gut shows considerable similarity to that of humans at the phylum level. Pigs also represent a useful in vivo model for investigation of transplanted human fecal microbiomes under certain conditions [29]. Furthermore, the pig gut is immunologically similar to that in humans since it contains both Peyer's patches and lymphocytes that participate in mucosal immunity. The predominant microbiota of humans and common laboratory animals typically contains *E. coli*, *Streptococci*, *Lactobacilli*, *Clostridium perfringens*, yeasts, and bacteroides [15, 30]. The upper gastrointestinal tract in humans and rabbit is similar and contains members of the Bacteroidetes and *Bifidobacterium* spp. (a probiotic microbe that is low to nonexistent in the porcine gut). Pigs are susceptible to many enteric pathogens affecting humans [28, 29]. Understanding the subtle microbiome distinctions can help choose the correct animal model for studies on absorption, bioavailability, nutrition, food transit, and gut-brain nexus within humans [15, 17, 31, 32].

The complex structure and functioning of the gastrointestinal tract lend itself to a multitude of gastrointestinal disorders [17, 18, 33–35]. Esophageal disorders (such as gastroesophageal reflux disease (GERD), infection, corrosion or rupture of veins, and motility disorders) lead to sore throat, difficulties in swallowing, and regurgitation. Investigation of the esophagus is performed using barium swallows, gastroscopy, endoscopy, and mucosal biopsy. Stomach disorders (such as inflammation caused by infection such as *Helicobacter pylori*, gastric ulceration, and peptic ulcers) result in indigestion, vomiting, malnutrition, or vomiting blood in chronic cases. Small intestine disorders (such as inflammation, peptic ulcers, malabsorption, and tumors) cause diarrhea, malnutrition, lethargy, and/or weight loss. Investigation of the small intestine is conducted by blood tests, diagnostic imaging with and without contrast, and small intestinal endoscopy. Large intestinal disorders (such as inflammation, functional colonic diseases, and colonic infections by enteropathogens) may cause bloody stool, constipation, fever, and/or abdominal pain. Investigation of the large intestine is carried out using diagnostic imaging, colonoscopy, and capsule endoscopy. Rectal disorders (such as inflammation, hemorrhoids, and anal cancer) result in anal pain while defecating, fresh blood in the stool, and constipation. Investigation of the rectum is performed through regular medical tests, rectal examination, and proctoscopy. Disorders of the liver, pancreas, gall bladder, and biliary tract (e.g., gallstones, inflammation, and cancer) adversely influence the secretion of digestive enzymes and gastrointestinal functions and are examined by laboratory tests to measure the levels of enzyme activities and specific organ function.

*1.2. Time Evolution of Ingestible Wireless Telemetric Capsules.* Historically, the basic concepts of endoscopy and its relevance in medical diagnostics were known to us for more than a century [36]. Rigid-wire endoscopy dates back to 1852 but was limited to the upper gastrointestinal tract. As summarized by Dr. Paul Swain (renowned gastroenterologist at the Royal London Hospital, London, U.K.), the potential benefits of imaging different sections of the gastrointestinal tract (via wireless capsule endoscopy) were recognized in parallel with the need for detecting its biophysical and biochemical signatures (via ingestible electronics) [37]. In 1954, Harold Hopkins invented the fibrescope (i.e., an optical unit composed of flexible fiberoptic bundles) to send optical images of the gut. This led to the invention of a fully flexible, digestive endoscope by Basil Hirschowitz in 1957. A collection of photographs of these pioneers and the endoscopic equipment from that particular era is nicely illustrated in a recent review article [10]. The same decade witnessed the invention of the first wireless, swallowable capsule (called the radio pill) to measure pH, core body temperature, and pressure of the gastrointestinal tract. In 1981, the first wireless video capsule was developed to capture images of the gastrointestinal tract wall, even though imaging technologies were rudimentary. Heidelberg Medical Inc. (Germany) developed a pH capsule to diagnose abnormal hydrochloric acid production in the stomach. Its tethered capsule (15.4 mm long by 7.1 mm wide) consisted of a radio frequency (RF) transmitter with an electrode and had a 6-hour lifetime. In 1997, capsule endoscopy was conducted for the first time on live and deceased pigs using capsule prototypes equipped with a light source, camera, video processing unit, battery, and transmitter [38].

By the early 2000s, low-power CMOS image sensors paved the way for incorporating video capture functions within wireless capsules [37]. There were significant improvements in the safety and performance of capsule technologies through adopting smart sensing devices, application specific integrated circuits (ASIC), improved white light optics, and adaptive signal processing [1, 8, 39]. One of the first specialized imaging capsules was the M2A (i.e., mouth-to-anus) and PillCam SB capsules developed by Given Imaging (Israel). The next-generation PillCam SB 2 capsule had wider-angle lenses, autoexposure feature, and a longer battery life. The esophagus-specific (PillCam ESo) and colon-specific (PillCam Colon) capsules were released with higher frame rates. Simultaneously, companies such as Olympus (in Japan), IntroMedic (South Korea), and Jinshan (China) developed their novel capsule endoscopy technologies. In 2003, the SmartPill became the first FDA-approved wireless capsule technology that integrated sensors for measuring pH, core body temperature, and pressure.

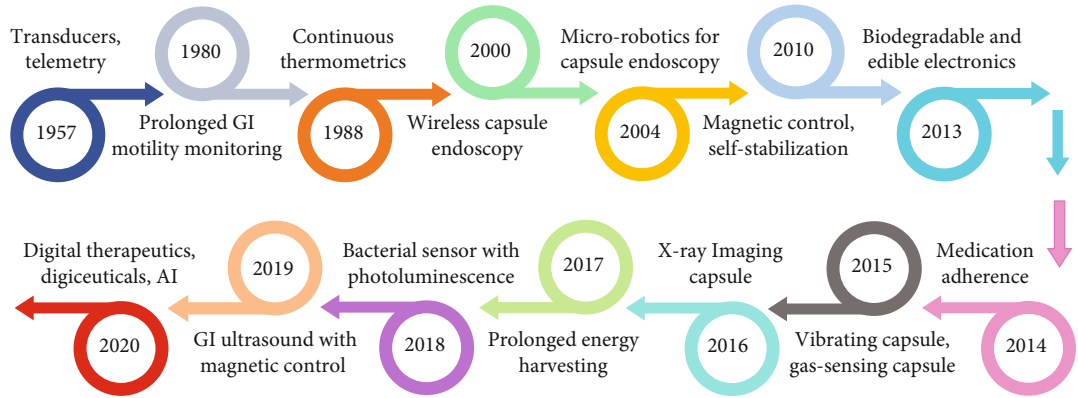

Figure 2: Timeline of ingestible capsule technologies. The evolution of ingestible capsules is illustrated here to show the important discoveries related to prolonged motility monitoring, continuous thermometrics, wireless capsule endoscopy, microrobotics, magnetic control and self-stabilization, biodegradable and edible electronics, medical adherence tracking, vibrating capsules, gas-sensing capsules, X-ray and ultrasound imaging, energy harvesting, bacterial sensing, digital therapeutics, digiceuticals, and artificial intelligence (AI).

Since 2004, there has been pioneering work in controlling the navigation and localization of capsule endoscopes within the gastrointestinal tract by employing magnetic fields, microrobotics, and self-stabilization techniques. Edible and bioresorbable electronics were explored to reinvent the capsule circuitry and other system components. In the past decade, new frontiers in gastrointestinal diagnostics and therapy have been explored, such as real-time gas-sensing, X-ray and ultrasound imaging, vibrating capsules, medical adherence, microbiome studies, digital therapeutics, digiceuticals, and artificial intelligence. A timeline of the abovementioned ingestible capsule technologies is depicted in Figure 2. From the clinical perspective, emphasis started to be placed on improving the quality and safety of endoscopic service with a patient-centered approach [40]. As such, specialized nursing groups were created within gastroenterology and endoscopy societies to improve the safety and comfort of patients, monitor the purchase and maintenance of endoscopic equipment, and provide support during patient hospitalization, recovery, and discharge [10].

*1.3. FDA Guidelines for Ingestible Wireless Telemetric Capsules.* The Food and Drug Administration (FDA) has classified "ingestible wireless telemetric capsule systems" as class II medical devices under 21CFR §876.1300, product code NEZ [41]. The manufacturers of ingestible capsules are required to have a Premarket Notification 510(k) which shows that the device is substantially equivalent to a product already in legal commercial distribution within the United States. The FDA Class II Special Controls Guidance provides directions on how to collect clinical information on various topics, such as the ease of capsule ingestion, intestinal transit time, diagnostic yield, any adverse events, and agreement amongst reviewers on the interpretation of images [41].

The FDA has identified six risks to health from this "generic gastroenterology and renal device": (i) biocompatibility, (ii) electrical and mechanical safety, (iii) intestinal obstruction or injury, (iv) functional reliability, (v) misinterpretation of recorded images, and (vi) RF-radiated power and electromagnetic compatibility [41]. The FDA guidance indicates that the testing on mechanical and structural integrity can be related to the battery life, field of view, depth of focus, and device exposure to pH levels and mechanical forces that are typically experienced during clinical studies [41]. For software-controlled medical devices, the FDA requires sufficient evidence of software functions and performance testing which is essentially determined by the "level of concern" (e.g., minor, moderate, or major) resulting from software failure [41, 42]. Patient labelling should have instructions on proper operation, monitoring, maintenance, reporting, and information on the risks and benefits. Some examples of items to be included in patient labelling are dietary restrictions, limitations on physical activity, safety features, use limits, possible symptoms (e.g., nausea, pain or vomiting), warnings, precautions, contraindications, and electromagnetic interference [41]. Lastly, in order to facilitate smooth licensing and approval, the FDA mentions that it considers the least burdensome approach for device manufacturers to comply with its guidance and address the identified issues [41].

## 2. Capsule Components for Gastrointestinal Sensing, Imaging, and Wireless Telemetry

Most ingestible capsules have a cylindrical form factor in the standard 000 capsule size (diameter = 11 mm and length = 26 mm). The ingestible electronics consists of miniscule

Table 2: Materials for next-generation ingestible capsules.

| Categories | Example materials |
| --- | --- |
| Insoluble materials | Agar, gelatin, cellulose acetate, ethyl cellulose, polyethylene, polyamide, polyvinylchloride, polyvinyl acetate |
| Soluble materials | Carboxymethyl cellulose, methylcellulose, hydroxypropyl cellulose, croscarmellose, hydrogels, hydroxyethyl cellulose, hypromellose, starch, sugars, polyvinyl alcohol, gums, alginates, polyacrylates |
| Biodegradable substrates | Cellulose nanofibril paper, polycaprolactone (PCL), poly lactic-glycolic acid (PLGA), rice paper, silk fibroin, poly (1,8- octanediol-co-citrate) (POC) |
| Flexible substrates | Chitin, gellan gum, pectin, collagen, sodium alginate, silk fibroin, polyesters |
| Fillers | Glucose, lactose, starch, oxides, carbonates, bicarbonates, sulfates, nitrates, magnesium silicate, NaCl, KCl, alkali metals phosphates |
| Plasticizing agents | Dibutyl sebacate, polyethylene glycol, polyethylene oxide, triacetin, triethyl citrate |
| Responsive materials | Enzyme-sensitive polymers (chitosan, starch), pH-sensitive polymers (polymethacrylates, enteric elastomer), oxygen-sensitive polymers, temperature-sensitive polymers (cyclododecane, methanesulfonic acid wax), moisture-responsive polyanhydrides |
| Foodstuff | Crackers, corn chips, fruit roll ups, gelatin, gummy, potato, pasta, rice paper, rice starch |
| Protective coatings | Diamond-like carbon (DLC), tetrahedral amorphous carbon (ta-C), gelatin, shellac |
| Biopigments | Melanins, chlorophylls, carotenoids, indigo, flavonoids, quinones, cytochromes, carotenes |
| Conductive elements | Copper, gold, silver, iron, platinum, magnesium, molybdenum, zinc, metal alloys, activated carbon, graphene; conductive polymers; ion-containing hydrogels; conductive paste |
| Thin films | Silicon nanomembranes, carbon, molybdenum, iron, tungsten, zinc, polymers |
| Semiconductors | Silicon, germanium, silicon germanium, indium-gallium-zinc oxide (IGZO), zinc oxide |
| Dielectrics | Silicon dioxide, silicon nitride, magnesium oxide, sucrose, albumin, spin-on-glass |
| Battery anodes | Magnesium, zinc, sodium, lithium, iron, iron oxide, tungsten, palladium, molybdenum, and alloys |
| Battery cathodes | Copper salts (chloride, bromide, iodide, sulfate), ferric salts (orthophosphate, pyrophosphate), vanadium oxide, manganese oxide |

*sensors* to detect physiological signals and capture images, *ASIC chips* to amplify and digitize sensors' output, and *wireless telemetry* to communicate the recorded data to an external receiver for data analytics and predictive diagnostics.

2.1. Ingestible Physiological Sensors, ASIC Chips, Optics, and Batteries. The physiological sensors and ASIC chips for ingestible electronics are generally fabricated in the standard silicon CMOS or CCD technology. For capsule video endoscopy, CMOS image sensors have faster readouts, smaller pixel size, and less power consumption, but suffer from noise susceptibility. On the other hand, CCD image sensors produce images of high quality and low-noise, but are inherently power hungry. Various sensors and ASIC chips can also be purchased directly from semiconductor manufacturers and distributors of commercial electronics and ASIC chips (e.g., Mouser, Arrow Electronics, Digi-Key Electronics, Future Electronics, Avnet, Rutronik, and Newark Electronics). Image compression techniques, such as compressed sensing (CS), are employed to dramatically improve the efficiency of image reconstruction from sparse signals using fewer samples than that required by the Nyquist-Shannon theorem [43]. A polymeric coating covers and protects the electronic and optical devices. Typically, button silver-oxide batteries provide the power, each having an output of 1.55 V and 20 mW. The silver-oxide battery has a silver(I) oxide cathode, zinc anode, and an electrolyte (sodium hydroxide or potassium hydroxide). The silver(I) ions are reduced to silver at the cathode, while zinc is oxidized to zinc(II) at the anode. The outer encapsulation material is chosen to make the capsule incompressible, nondisintegrable, and nondegradable during its transit through the gastrointestinal tract. Next-generation ingestible capsules are harnessing new materials from nature to redesign various components such as coatings, substrates, semiconductors, conductors, dielectrics, and batteries (Table 2).

2.2. Wireless Telemetry, Transmission Frequency Bands, and RF Transmitters. The design of wireless telemetry for ingestible electronics is dependent on various parameters such as transmission frequency, frame rate, data rate, pixel resolution, specific absorption rate (SAR) limit, and power consumption [44]. A comprehensive table of RF band allocations within Australia and the United States was recently published [45]. The ISM radio frequency bands are reserved for industrial, scientific, and medical (ISM) purposes (433.05-434.79 MHz for Europe and 902-928 MHz for the U.S.) [46]. While the optimum transmission frequency from an ingestible capsule in the human body is 600-800 MHz, transmitting signals at lower frequencies (e.g., 32 MHz) requires smaller real estate, reduced power consumption, and longer battery life [46]. The common RF communication protocols suitable for ingestible capsules are Bluetooth Low Energy (BLE), ZigBee, and WiFi [45]. Other protocols, such as UWB, ANT, Thread, LoRa MICS, IrDA, RFID, and NFC, also have potential use within this field. WiFi is designed to wirelessly connect consumer electronic devices (e.g., phones and laptops) to the network and hence requires large amount of power [45]. Zigbee and LoRa are better suited for embedded systems with low power consumption but have relatively smaller range [45]. BLE implemented with Bluetooth 5.0 appears to be best suited for low power transceivers and consumer applications with a midpower consumption [45].

After choosing a suitable transmission frequency, the design and test of the RF transmitter is a critical aspect of wireless telemetry. The RF transmitter consumes a significant portion of the power to overcome the signal attenuation from human body tissues, which can be greater than 23.4 to 24.4 dB (based on estimates of a 916.5 MHz electromagnetic source located within 15 cm of human tissue) [47]. As a reference, the dielectric parameters of the human gastrointestinal tract are as follows: relative permittivity $\varepsilon_r = 67.2$ and conductivity $\sigma = 1.01 \, S \cdot m^{-1}$ [48]. To study the electromagnetic radiation from RF transmitters, numerical modelling simulations and tissue phantoms are preferred [49]. Helical antenna are best suited for ingestible capsules because they provide omnidirectional radiation distribution, circular polarization, and consistent bandwidth across different tissues [50, 51]. The PillCam SB has used helix antenna (width = 8 mm, diameter = 5 mm) with seven turns within its internal structure, which takes up considerable real estate within the capsule. As an alternative, the Sonopill uses a conformal helix antenna that wraps around the exterior shell of the capsule to conserve interior space.

Modelling the channel attenuation of RF signals between the endoscopy capsule systems and the external receivers has been previously attempted for the ISM 2.4 GHz band and the 402 to 405 MHz Medical Implant Communication Service (MICS) band [52, 53]. The ISM 2.4 GHz band supports high rate of data transfer that is appropriate for several wireless applications but leads to higher chances of interference from other wireless devices, greater attenuation, and larger power consumption [53]. On the other hand, the 402 to 405 MHz MICS band has reduced attenuation, lower interference from other wireless devices, and requires low power but does not support high rate of data transfer [53]. Previous studies have computed the path loss (PL) and specific absorption rate (SAR) using homogeneous or heterogeneous tissues and often complemented by simulation results from 3D electromagnetic solvers, such as the SEMCAD-X package (SPEAG, Zurich, Switzerland) [52, 53]. The conductivity and dielectric constant varies for the different tissues, and earlier studies have employed models of the muscle tissue, skin tissue, and esophagus tissue to characterize these variations [53]. The path loss is directly proportional to medium's conductivity and inversely proportional to the square root of medium's permittivity [52]. The path loss also increases when the separation between the antennas is increased or when the antennas are misaligned [52]. To give some quantitative insights on the attenuation of the RF signals through the human tissue, one study modelled a capsule system with a spiral antenna [53]. At 402 MHz, the reflection coefficient was between -5.5 dB and -7 dB with the -5 dB bandwidth around 45 MHz [53]. The path loss ranged from 58 dB in the muscle tissues to 61 dB in the esophagus tissue using aligned antennas separated by a distance of 250 mm, which was considered satisfactory for effective data communication in capsule endoscopy systems [53]. For the same setup, the peak spatial SAR was around 2.2 mW/kg, which was considerably lower than the 2 W/kg limit set by the International Commission on Non-Ionizing Radiation Protection (ICNIRP) [53].

Some examples of commercial wireless sensor nodes (available as one complete board) are the Mica2DOT (by Crossbow), T-node (by SOWNet), Sensium (by Toumaz), and Tmote Sky (by Sentilla) [54]. The Mica2DOT sensor node (dimensions: $58 \times 32 \times 7 \, mm^3$, weight: 18 g) uses the ATmega128L microcontroller (8 MHz, 4 k SRAM, and 128 k flash memory). It operates at 38.4 kb/s data rate and multiple frequency bands (868/916 MHz and 433 MHz) while consuming 135 mW power at 3.3 V supply. The T-node is comparable to the Mica2DOT in its physical dimension (23 mm diameter), operating frequencies, and data rates (50 kb/s). The T-node uses a separate chip for the 10-bit ADC, 128 k flash memory, and 4 k SRAM. However, the data transmission rates of both these sensor nodes are low for reliable video capsule endoscopy. To accommodate faster data rates, a high frequency telemetry system (1.2 GHz and 20 MHz bandwidth, 20 Mb/s data rate) has been demonstrated, even though the 1.2 GHz frequency band is unavailable for medical device applications [55]. With rising demand for consumer electronics, a number of low-power Bluetooth System-on-Chip (SoCs) chipsets are now available, such as CC2540 and CC2650MODA by Texas Instruments, nRF1822 by Nordic Semiconductor, and BGM11S by Silicon Labs [45].

*2.3. Examples of Sensors, Devices, and Circuits for Ingestible Capsules.* Below are some examples of sensors, devices, and circuits within experimental capsules to measure the gut pH, core body temperature, gastrointestinal motility, and medical adherence.

(a) A radiotelemetry capsule was designed for the assessment of pH and core body temperature within the gastrointestinal tract [56]. The capsule had integrated a pH ISFET device and a silicon temperature sensor. The ASIC chip was fabricated in a $0.6 \, \mu m$ CMOS process. The chip included a sensor interface (six operational amplifiers and a multiplexer), timer (with 32 kHz RC relaxation oscillator), and a system scheduler (finite state machine with a 10-bit ADC). A double-sided printed circuit board (PCB) was used to assemble the ASIC chip, two SR48 silver oxide batteries, a magnet for telemetry, and RF transmitter (operating at 433.92 MHz). The capsule ($36 \times 12$ mm, 8 g) had a power consumption of 15.5 mW in the active mode. The capsule was operational for 42 hours during *in vitro* testing in artificial gastric and intestinal solutions

(b) A wireless tablet-shaped ingestible capsule was developed to monitor the core body temperature

[57]. Power was generated by a battery with gold and magnesium electrodes, and activated by gastric acid. The battery voltage output was boosted and stored in a 220 μF ceramic capacitor. The ASIC chip was fabricated in 0.6 μm CMOS technology and included a temperature sensor, microcontroller, a resonant circuit, sequencer, coder, modulator, and transmission coil. Data transmission was carried out with binary phase-shift keying (BPSK) modulation at 13.56 MHz, followed by wireless telemetry using magnetic field coupling

(c) A wireless capsule was fabricated in 0.18 μm CMOS technology to identify gastrointestinal motility disorders. The capsule integrated three sensors (pH, temperature, and pressure), low power ASIC, batteries, RF transceiver, and Helix antenna [58]. The external data recorder had a RF transceiver with antenna, batteries, multimedia card, and migrating motor complex. A computer workstation performed the data preprocessing and analytics. The sensing units were a thermistor, pH-sensing ISFET, and a C29 AKTIV pressure sensor. The ASIC chip had an 8-bit microcontroller, clock, and power management modules and 18-bit ADC, communication processing unit, and memory. The RF transceiver used FSK modulation to transmit and receive data at 9.6 Kb/s with a 434 MHz center frequency. The chip (5 mm × 5 mm) used a 3.3 V supply voltage, 2-8 MHz clock frequency, and 300 μA current drainage at 2 MHz. The capsule was tested on three human volunteers and 20 cases of human experiments in multiple hospitals. While its data analytics could not distinguish patients with astriction or diarrhea, its clustering algorithm could correctly recognize 83.3% of patients with absent gastrointestinal motility

(d) As an alternative to batteries for powering the capsule electronics, wireless power transmission was demonstrated in a radiotelemetry capsule robot [59]. Two separate communication bands were used—a lower frequency for power transmission (13.56 MHz) and a higher frequency for data carriers (433.92 MHz). The capsule robot (10 mm × 25 mm) had pressure and temperature sensors, FSK modulation transceiver (data rate of 112 kbps, 433.92 MHz located in the ISM band), class-E power amplifier and oscillator, 16-bit ADC, timer module, flash storage module, power management module, and signal-processing microcontroller [59]. The receiving coil within the capsule induced a voltage by inductive link with the transmitter to receive up to 280 mW of wireless power at 13.56 MHz. The sampling rate of the capsule robot was 0.6 Hz (i.e., 1 sample per 1.67 seconds), which was sufficient to capture the gastrointestinal motility signals in humans (that are from 1 to 18 cycle per minute). A data recorder was worn around patient's waist and used a multimedia card to store incoming data at 433.92 MHz and later pass it to a workstation

(e) The Proteus Digital Health platform consisted of an ingestible sensor and wearable patch to track poor medication adherence, especially in patients who self-administer their prescriptive drugs at home [60]. Its ingestible sensor (1 mm × 1 mm × 0.3 mm, CMOS integrated circuit) was activated by gastric fluids to transmit signals at 10-30 kHz and 200-800 bits/sec using BPSK modulation. The data pod within the wearable patch had power circuits, low noise amplifier, custom system-on-chip (SoC), LEDs, accelerometer, temperature sensor, Bluetooth radio, ceramic radio, and coin cell battery. The data pod received and decoded the signals from the ingestible sensor, in addition to monitoring the step count, body angle, and heart rate. The data stored in the data pod's flash memory was transferred to a mobile device using Bluetooth Low Energy (BLE) protocols. The ingestible sensor had a high detection accuracy of over 98% in 1,200 unique ingestions recorded across 14 individuals [60].

## 3. Capsule Navigation and Localization in the Gastrointestinal Tract, Capsule Placement Devices

Most wireless ingestible capsules rely solely on the natural peristalsis movement to navigate the digestive tract. Capsule movement using peristalsis is satisfactory when capsule's orientation, velocity, and resident time within each gastrointestinal region are less important. However, guided capsule movement using microrobotics and localization techniques (e.g., magnetic actuation and electrical stimulation) is desired for taking a closer view of specific regions of the gastrointestinal tract or controlling the resident time in a specific region [6, 8].

*3.1. Navigation by Magnetic Actuation and Robotic Manipulators.* Magnetic actuation employs electromagnetic coils or permanent magnets to generate and control the magnetic fields for capsule navigation and localization within the gastrointestinal tract [61–63]. Electromagnetic systems modulate the magnetic field strengths by changing the coil current, though high coil currents lead to Joule heating and safety concerns [61]. Compared to electromagnetic coils, permanent magnets are cost-effective and have a compact form factor. In conjunction with handheld devices or robotic guiding systems, permanent magnets within a capsule can apply desired forces and torque to the capsule, thereby facilitating better orientation during imaging and faster transit speeds through the gastrointestinal tract. Magnetic steering systems may have three types of propulsion: translational, rolling, or helical. Translational steering systems use nonrotational magnetic fields, while rolling and helical propulsion systems require rotating magnetic fields. Today, magnetic steering systems can have five degrees of freedom (DOF) (i.e., three translational DOF and two rotational DOF) as demonstrated by a team of researchers building a wireless capsule gastroscope with a singular permanent magnet system [64]. The team used an external robotic magnetic manipulator with 6-DOF to maneuver the capsule. A magnetic force controlled capsule's position and translation while a magnetic torque controlled capsule's rotation and heading. The 5-DOF manipulation of the magnetic capsule was tested in an *in vitro* environment simulating a fluid-filled stomach [64]. On the commercial front, Stereotaxis has a number of products involving the "robotic magnetic navigation" (RMN) technology. Stereotaxis Niobe RMN system uses two robotically controlled magnets to adjust the magnetic field and precisely steer a catheter (with a magnet at its tip) inside a patient. The Stereotaxis Genesis RMN system uses smaller magnets (compared to the Niobe system) with better flexibility of the robotic arms, improved patient safety and outcomes, and unprecedented responsiveness to physician control during procedures.

*3.2. Navigation by Electrical Stimulation.* Besides magnetic actuation, electrical stimulation has been investigated for guiding the locomotion of ingestible capsules through the gastrointestinal tract [65]. In one study, a radio-controlled electrostimulation capsule (12 mm × 30 mm) was constructed having a RF receiver, decoder system, three 3 V lithium batteries, and two peripheral nervous system electrodes [66]. The electrodes reached a maximum voltage of 18 V. An external controller was used to change the amplitude and period of electrical stimulations, which directly affected capsule's velocity. The electrical stimulation of intestinal walls allowed for capsular propulsion of up to 3.4 mm/s in the aboral direction. *In vivo* tests using porcine intestinal tracts yielded a correlation between capsule's speed and voltage amplitude, with the optimal voltage amplitude between 6 V and 9 V. In another study, a predictive mathematical model for the electrostimulated capsular motion in the small intestine was built using Stokes' drag equation and regression analysis [67]. The model considered friction and contraction force as the primary forces acting on a capsule during its transit. The model simulations provided the optimal shape and size of the capsule. Thereafter, actual capsules were constructed having similar internal features and having two sets of electrodes to allow for movement in the oral and aboral directions. The capsules were tested in 72 experiments with stimulations having a constant voltage and duration (i.e., 6 V, 5 millisecond pulse duration) and various frequencies (i.e., 10 Hz, 20 Hz, and 40 Hz). The experiments yielded an average forward velocity of $2.91 \pm 0.99$ mm/s and an average backward velocity of $2.23 \pm 0.78$ mm/s for the capsules. Unlike magnetic actuation, the safety and efficacy of electrical stimulation techniques still needs further validation through clinical studies to realize its anticipated commercial success.

*3.3. Capsule Placement Devices.* Capsule placement devices have been developed for individuals who are candidates for video capsule endoscopy but have difficulty swallowing the video capsule or passing it through the pylorus. Such an individual may have known or suspected motility disorders or anatomical abnormalities (e.g., dysphagia, gastroparesis, or large hiatal hernias) that could result in stalled capsules if directly swallowed through the mouth. The capsule placement device is used to facilitate direct endoscopic placement of video capsules into the stomach or duodenum while posing minimal intubation risks to the esophagus.

As an example, the AdvanCE capsule endoscopy delivery device by STERIS has a sheath (diameter = 2.5 mm, length = 180 cm) and a clear capsule cup for atraumatic capsule deployment into the stomach or duodenum. The capsule cup firmly holds the video capsule during endoscopic intubation of the esophagus, and the capsule is then released by pulling an outer handle upon reaching the target GI section. The AdvanCE device is compatible with most video capsules (diameter = 10.5 mm to 11.5 mm, length = 23.5 mm to 26.5 mm), such as those manufactured by Olympus, Given Imaging, and IntroMedic. By directly placing the capsule within the stomach, the battery life spent passing through the esophagus or stomach is saved.

## 4. Commercialized Capsules for Physiological Status Monitoring

Below are examples of commercial ingestible capsules to monitor the core body temperature, gut pH and acid reflux, and intestinal gases. Table 3 lists the companies for physiological status monitoring, sample products, dimensions, features, specifications, and clinical applications.

*4.1. Core Body Temperature Monitoring.* Maintaining the core body temperature consumes approximately 40% of body's energy expenditure [68]. The core body temperature is not fixed but can fluctuate depending on the amount of food or drinks consumed, environmental factors, and physical fitness of the individual [69, 70]. Measurements of the core body temperature can be made invasively through the oral, esophageal, or rectal cavity. Oral and esophageal readings are not often accurate, while rectal temperatures are inconvenient. The core body temperature can also be predicted from the skin temperature and heat flux measurements at different body sites, but the method is not reliable in outdoor settings. As an alternative to invasive and indirect temperature monitoring methods, ingestible capsules are capable of unobtrusively tracking the internal body temperature in mobile users, such as athletes, fire fighters, emergency care responders, and soldiers [71–73]. Capsule-based monitoring of the basal body temperature while sleeping helps to manage the inner-body clock and sleep-wake rhythm [72, 74, 75]. Continuous, real-time monitoring of the core body temperature can assist people to stay within safe levels of hyperthermia while preventing overexertion, heat stress, and heat stroke. Some examples of commercial capsules for core body temperature monitoring are described below:

(a) The CorTemp by HQ Inc. is an ingestible body thermometer pill that is FDA cleared and registered as a single-use device (Figure 3). The CorTemp pill has a silicone-coated sensor with a quartz crystal oscillator, circuit board, battery, and communication module. In the gastrointestinal tract, its quartz crystal

TABLE 3: Commercialized ingestible capsules for GI physiological status monitoring.

| Company | Sample products | Size ($D \times L$), weight | Features and technical specifications | Clinical applications |
|---|---|---|---|---|
| HQ Inc. | CorTemp thermometer pill, CorTemp recorder | 10.7 mm × 22.4 mm, 2.8 g | Microbattery, quartz crystal, communication coil and circuit board; 262 kHz or 300 kHz | Monitor core body temperature to prevent heat stress and heat illnesses |
| Philip Respironics | VitalSense capsule, dermal patch, XHR sensor | 8.6 mm × 23 mm, 1.6 g | 25 °C to 50 °C temperature range, 240 hrs battery life, 1 meter transmission range | Monitor core body temperature, dermal temperature, heart rate, and respiration rate |
| BodyCap | eCelsius medical capsule, performance capsule | 8.9 mm × 17.7 mm, 1.7 g | Accuracy of ±0.2°C in the range 25-45°C, 20 days battery life, 1 meter range | Sports performance; medical/industrial health monitoring; severe heat or cold environments |
| SmartPill Corporation | SmartPill wireless motility capsule | 13 mm × 26 mm, 4.5 g | 434.2 MHz, silver oxide batteries, 25°C-49°C, 0.05-9.0 pH range, 0-350 mmHg pressure range | Used for capsule motility procedures; alternative to scintigraphy and radio-opaque markers |
| AnX Robotica | VibraBot capsule | 11.8 mm × 26.7 mm, 4.5 ± 0.5 g | Adjustable vibration frequency, 180 minutes continuous vibration, smartphone control | Nonpharmacological constipation relief, mechanically stimulate peristalsis in large intestine |
| Vibrant Gastro | Vibrant Capsule | 11.3 mm × 24.2 mm, N/A | Flat motor, batteries, electronic card, algorithm-based control, adjustable vibration frequencies | Chemical-free constipation treatment capsule, relief from chronic idiopathic constipation |
| Atmo Biosciences | Atmo gas-sensing capsule | 11 mm × 20 mm, N/A | Measures hydrogen, oxygen, carbon dioxide, methane; data sent every 5 min for 72 hrs | Measure gas production, onset of food fermentation, and gut microbiome activity |
| Proteus Discover, Otsuka Pharma | Proteus Digital Health, Abilify MyCite, ProteusCloud | N/A | Ingestible pill with biosensors, wearable patch, medication tracker app, single dose resolution | Track medical adherence for drug therapeutics; check if oral medications were ingested (e.g., for tuberculosis, schizophrenia and oncology) |

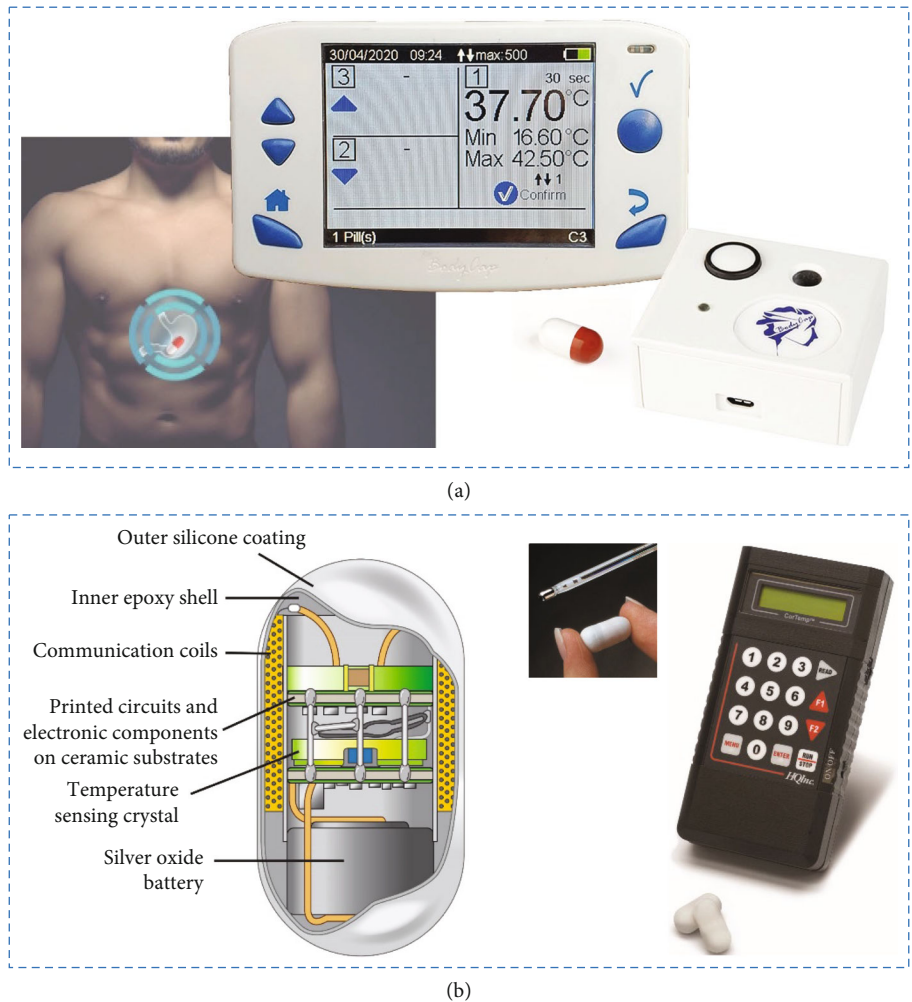

Figure 3: Continuous core body temperature monitoring capsules. (a) The BodyCap e-Celsius Core Body Temperature Ingestible Capsule measures the gastrointestinal temperature and transmits the data to an external e-Viewer Performance Monitor. (b) The CorTemp Ingestible Core Body Temperature Sensor transmits information on the internal body temperature to the CorTemp Data Recorder worn externally by the user. The images were reproduced with permission from BodyCap (a) and HQ Inc. (b).

oscillates at a frequency relative to the core body temperature. This creates a corresponding magnetic flux, and the information is transmitted to the CorTemp Data Recorder (262 kHz to 300 kHz). The pill has been applied to the detection and prevention of heat stress and heat-related illnesses for sports, research, industry, and medicine

(b) The eCelsius by BodyCap is an ingestible electronic capsule for the continuous monitoring and wireless transmission of core body temperatures (Figure 3). The eCelsius Medical has been used to diagnose a patient's febrile state for applications in sleep disorders, oncology, and infectious diseases. The eCelsius Performance is designed for athletes and swimmers who need continuous core temperature monitoring without any constraints on the subject. Their eViewer monitor manages the data from up to 3 capsules within a 1-3-meter range. Some other eCelsius applications include the early detection of peak fever during infections and chemotherapy and tracking the circadian rhythm of core body temperatures for sleep analysis

(c) MyTemp BV has developed a battery-free, core body temperature monitoring capsule. It uses self-induction to power the capsule and transmit the gathered data to a belt worn on the waist. The MyTemp capsule has been used to detect overheating of athletes' core and to monitor animal heat stress in equestrian sports

*4.2. pH Monitoring and Gastroesophageal Reflux Testing.* Ingestible wireless pH testing capsules are being used to monitor pH levels over time for improved gastrointestinal diagnostics. One application area is esophageal pH testing, which is important for the diagnosis of gastroesophageal reflux disease (i.e., GERD) where a person suffers from frequent acid reflux. GERD has been traditionally diagnosed by a nasopharyngeal-wired pH monitoring system that causes patient discomfort. After swallowing the pH monitoring capsules, the pH profiles can be analyzed for four gastrointestinal landmarks: ingestion, pylorus, ileocecal valve, and excretion [76]. Ingestion is characterized by a rapid temperature rise and a rapid pH drop (>3 pH values), pylorus is characterized by a rapid pH rise (>3 pH values), ileocecal valve is noted by a rapid pH drop to ≤6.5, and excretion is noted by a rapid temperature drop [76]. Below are some examples of commercial ingestible wireless pH testing capsules for the gastrointestinal tract.

(a) The Bravo Reflex Testing System is an ambulatory pH test and primarily assists in the diagnosis of GERD symptoms. The esophageal pH is measured by an antimony pH electrode every 6 seconds over a period of 96 hours. This extended time period of pH monitoring increases the chances to detect reflux events and accurately determine the associated symptoms

(b) Jinshan Science and Technology has developed a wireless pH capsule system to record up to 96 hours of data. The wireless pH capsule is designed to be small in size ($6.0 \times 5.5 \times 26.5$ mm) and weight (1.4 g). A conveyor is used to deliver the wireless pH capsule to the lower esophageal sphincter (LES), followed by vacuum suction to attach the capsule to the mucosa. The pH sampling is fast (i.e., every 3 seconds), and the range of data transmission is sufficiently large (i.e., within 3 meters) for patient comfort. After traversing the gastrointestinal tract, the wireless pH capsule is naturally excreted from the body. An accompanying recorder provides automatic pairing with the pH capsule, double data protection, and improved user operability. Its data analysis software has options to autodetect events to reduce editing time, perform statistical calculations, and generate customized reports

(c) The SmartPill capsule endoscope was designed for the evaluation of gastric emptying and the measurements of transit time through the small bowel and colon. The standard techniques of studying gastric emptying by whole gut transit scintigraphy (WGTS) and radio-opaque markers require long-time durations and provide information only about colonic transit. The SmartPill capsule has integrated sensors to transmit data on the pH, temperature, and pressure in the gastrointestinal tract at regular intervals. The SmartPill MotilGI software helps to estimate different time parameters which have relevance to the gastrointestinal functioning. For instance, the whole gut transit time is calculated as the time between capsule's ingestion and its exit from the body (denoted by a sudden drop in temperature or loss of signal). The software also calculates the gastric motility index and gastric contractions per minute. The gastric residence time is calculated from capsule's ingestion to the sudden pH rise (i.e., >3 pH units from the acidic gastric region to the alkaline duodenum). The transit times through the ileum, cecum, and colon are calculated from the raw data related to pH, temperature, and pressure at various gastrointestinal sections [77, 78]. A clinical study used the SmartPill capsule to show that diabetic individuals with gastroparesis exhibited delayed colonic transit which correlated with delayed gastric emptying [79].

*4.3. Sensing of Intestinal Gases.* The process of carbohydrate digestion breaks down the poly- and disaccharides into absorbable monosaccharides which is followed by carbohydrate absorption in the intestine [30]. The unabsorbed carbohydrate reaches the colon where it is fermented by the bacteria [80, 81]. This bacterial fermentation can lead to excess gas production and other discomforts (e.g., bloating, diarrhea, abdominal distension, and irritable bowel syndrome (IBS)) [82]. Apart from the metabolic activity of intestinal bacteria with unabsorbed food content, various chemical conversions and enzymatic perturbations can produce different gases in the gastrointestinal tract, such as hydrogen, carbon dioxide, oxygen, nitrogen, and methane [83]. The malabsorption of various carbohydrates and small intestine bacterial overgrowth are commonly monitored from end-expiratory breath samples for levels of hydrogen or carbon dioxide [84]. Breath tests can also reveal elevated levels of methane in the gastrointestinal tract, suggesting higher colonization of methanogens (e.g., *Methanobrevibacter smithii*) in the gut who scavenge the hydrogen and produce methane as a by-product [85, 86]. The presence of both hydrogen and methane in the breath of individuals is correlated to higher body mass index and percent body fat [86]. However, studies show that hydrogen breath tests can give divergent or false-negative results due to several factors such as the presence of oral bacterial flora, nonhydrogen-producing bacteria, gastrointestinal motor disorders, or improper adherence to carbohydrate diet [35, 84]. Furthermore, the lack of accurate diagnosis in regard to the point of origin of the gas and very low signal-to-noise ratio limits the conclusions that can be drawn from the hydrogen breath test.

With the inherent deficiencies of breath tests and the inconvenient nature of other techniques (e.g., tube insertions, flatus analysis, and whole-body calorimetry), indirect measurements of gas production by *in vitro* fecal fermentation systems using colonic bacteria has drawn attention from experts [87]. Here, an oxygen-free bacterial incubation environment is created, and fecal inocula is cultured with nutrients on a fiber substrate. The gas released is absorbed in solid phase microextraction fibers and assessed for its composition using gas chromatography and mass spectrometry (GC-MS). To circumvent the offline gas analysis, portable gas-sensing capsules have been investigated to measure the intestinal gas species from *in vitro* fecal fermentation by employing electrochemical, calorimetric, or optical sensing modalities [87].

Wireless capsules have been demonstrated for the measurement of intestinal gas profiles *in vivo* and the discovery of biomarkers related to the gut microbiota [35]. These studies can potentially help to formulate better diets which positively influence the gut microbiota and colon health [81, 83]. A human pilot trial was conducted to test the efficacy of ingestible electronic capsules that monitor different intestinal gases (i.e., oxygen, hydrogen, and carbon dioxide) [88]. The capsule employed semiconducting metal oxide-based sensors and thermal conductivity sensors to measure the gas concentrations. The polyethylene capsule shell housed the gas sensors, temperature sensor, microcontroller, transmission system (433 MHz), and silver oxide batteries. Ultrasound was used to evaluate the physical location of the capsule and the transit time within different sections of the gut. In their study, four healthy volunteers were recruited, and subjected to high-fiber and low-fiber diets. The results showed that the oxygen-equivalent concentration profile was an accurate marker for tracking the physical location of the capsule, while hydrogen gas profiles represented the food fermentation patterns in the different sections of the gut. With high-fiber diet, there was only a small increase in hydrogen and carbon dioxide within the colon, which are indicative of minimal colonic fermentation. With low-fiber diet, there was an increase in colonic hydrogen and carbon dioxide with a significant shift in the microbiota as confirmed from metabolomics analysis of fecal samples. The team founded the Atmo Biosciences to commercialize their capsules (Figure 4) that can measure the intestinal gases at the source and identify digestive issues using clinical analysis and predictive algorithms [1, 35, 87, 89–92].

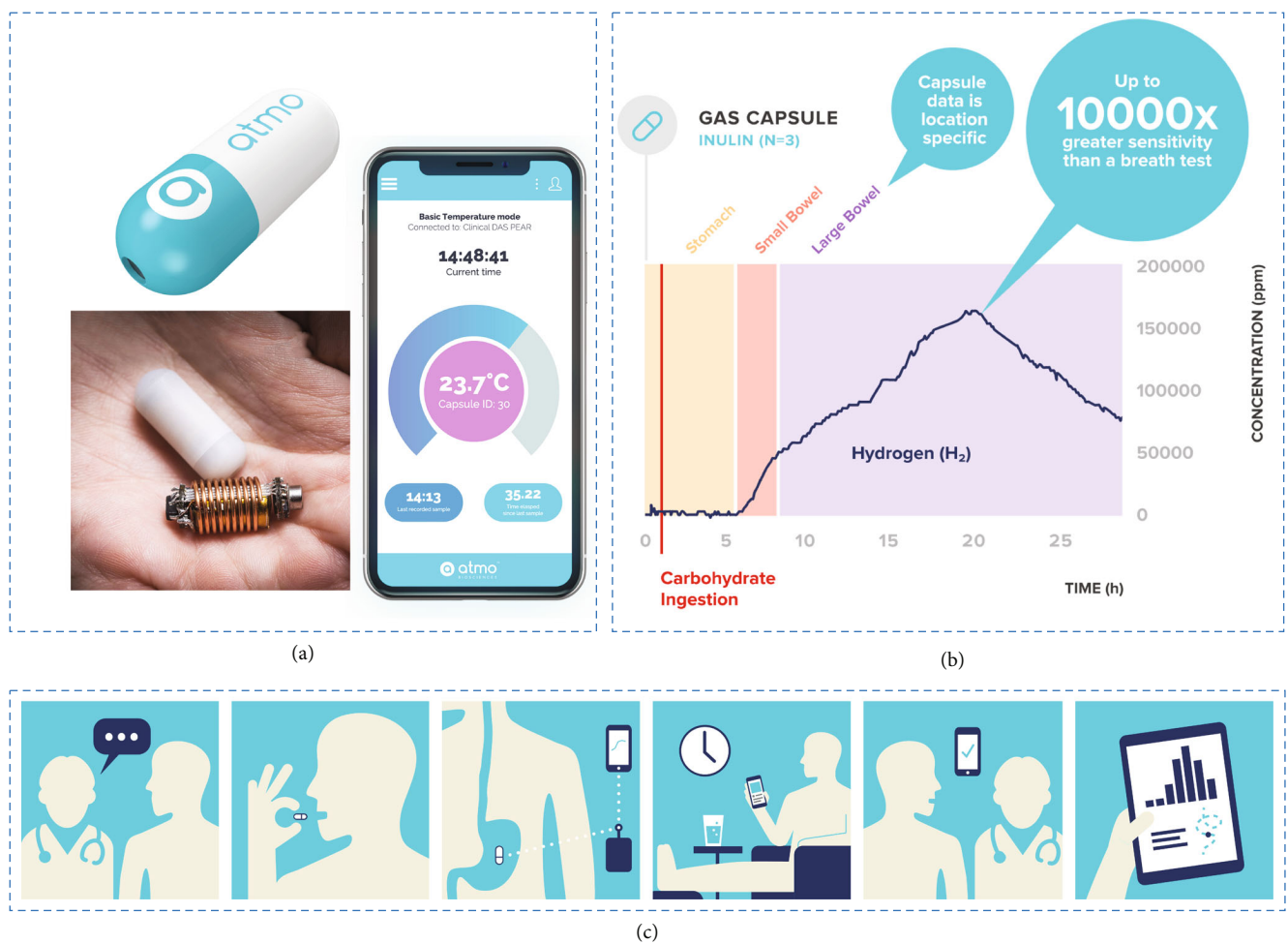

FIGURE 4: Gas-sensing capsules by Atmo Biosciences. (a) The Atmo Gas Capsule detects gases in the human gastrointestinal tract at real time for the diagnosis and targeted treatment of gut disorders. (b) The Atmo gas capsule has a higher signal to noise ratio compared to breath tests as the gas concentrations are measured at the site of production. (c) The patient can take the Atmo gas capsule at home, and the data is continuously transmitted to the physician through a smartphone app for analysis. The images were reproduced with permission from Atmo Biosciences.

## 5. Commercialized Capsules for Mechanical Vibrations and Bioelectric Neuromodulation

Below are some examples of commercial vibrating capsules for constipation relief and electroceutical capsules for obesity and hormonal release.

*5.1. Vibrating Capsules.* Functional constipation is classified as normal-transit, slow-transit, or outlet-transit types and is generally manifested by abdominal discomfort, stomach aches, and infrequent or painful defecation [93]. Current treatment options for chronic constipation include fiber supplements, stool softeners, osmotic and stimulant laxatives, lubricants, prokinetics, guanylate cyclase-C agonists (GC-C), and surgical management. The sustained effectiveness of the current treatment options for chronic constipation is not guaranteed due to their high recurrence rate, poor patient compliance, and treatment side effects. To improve the consistency of treatment for the aforementioned conditions, vibrating capsule technologies have been developed. Examples of commercial vibrating capsules for the treatment of functional or chronic constipation are as follows:

(a) Vibrant Capsule by Vibrant Ltd. is a nonpharmacological device to provide relief from constipation and restore the complete spontaneous bowel movements. As a chemical-free alternative to current treatment options for chronic constipation, the Vibrant Capsule generates mechanical vibrations in the wall of the large intestine to induce bowel movements and augment the normal circadian rhythm [93]. The Vibrant Base Unit activates the capsule using an electromagnetic signal. Different vibration modes can be programmed to regulate when and how to vibrate the capsule. The Vibrant Capsule has been tested on patients with Chronic Idiopathic Constipation and Irritable Bowel Syndrome with Constipation. In a study to evaluate its safety and efficacy, it was found that the vibrating capsule induced a significant increase in bowel movements per week in 23 out of 26 patients with no serious adverse events [94]. The team established safety by quantifying key parameters related to pathological anatomy, physiology, stool, blood, and CT scans [94]. A second study saw evidence of faster colonic transit time with the Vibrant Capsule treatment and mentioned that the vibration parameters require further optimization to accelerate the colonic transit in patients with functional constipation [95].

(b) The VibraBot by AnX Robotica is designed to provide nonpharmacological relief from constipation. One study tested the VibraBot capsule on the defecation frequency of beagle dogs [96]. An external configuration device enabled by a smartphone was used to communicate with the VibraBot capsule. Once the capsule reached the abdominal cavity, it was activated to generate abrupt vibrations at low, moderate, or high frequencies (with a frequency ratio of $0.5:0.8:1.0$, respectively). The VibraBot capsule produced mechanical vibration to stimulate peristalsis in the large intestine. Their results indicated that after capsule ingestion, there was a significant increase in the daily defecation frequency and a decrease in the capsule evaluation time

5.2. Electrical Stimulation Therapy and Bioelectric Neuromodulation. Besides gastrointestinal motility disorders, a class of electroceutical devices are being investigated as therapeutic tool for obesity, hormone release, and neural signaling [8, 97, 98]. Bioelectric neuromodulation (where electrical stimulation is given at target sites along the gastrointestinal tract) has gained significant interest as a method to relieve gastrointestinal disorders, even though the present outcomes are variable and the mechanisms of action are unclear [11]. Bioelectric neuromodulation may stimulate the vagus nerve, thoracolumbar connections, and sacral nerves to inhibit gastrointestinal inflammation, treat feacal incontinence, or block signals for feeding to reduce appetite and treat morbid obesity [11, 99, 100]. While there are no commercial examples of ingestible capsules for the electrical stimulation of the gastrointestinal tract, there are two precedents as implantable bioelectric neuromodulators for this purpose. In 2015, the VBLOC vagal blocking therapy by EnteroMedics Inc. was approved by FDA for the treatment of obesity. The EnteroMedics VBLOC therapy involved using its implantable pacemaker-like device, Maestro Rechargeable System, to intermittently block the vagal nerve pathway to reduce hunger and generate earlier feelings of fullness. This helps to promote a safe, healthy, chemical-free, and sustainable weight loss alternative to bariatric surgery (which has high surgical costs and increased rate of frequent complications) [101–103]. A ReCharge Study to evaluate the safety and efficacy of the EnteroMedics Maestro Rechargeable System found that VBLOC-treated patients reached 24.4% excess weight loss at 12 months [11].

## 6. Commercialized Capsules for Tracking Medication Adherence

One emerging application of ingestible capsules is to monitor and document the adherence to oral medications. The common reasons cited by patients for not taking their medications correctly can be forgetfulness, mental disorders, emotional factors, dearth of information, poor communication with caregivers, and deliberate intent to omit doses [104]. Poor medication adherence can result in treatment failure, multidrug resistance, and disease transmission to the community. This can also lead to frequent sickness, doctor visits, and hospitalizations. The two current standards of drug administration are *directly observed therapy* and *self-administered therapy*. Surveillance technologies can help monitor medical adherence, such as video recording by smartphones and manual time-register of the opening of pill bottles. However, the current methods do not ensure that all the required medications are actually ingested regularly at scheduled times. The clinicians also do not have rapid access to the medical adherence information from their patients. Connected medical devices are gaining attention for the remote tracking of drug management and patient outcomes. As an example of digital healthcare, a smart autoinjector, AdhereIT by Noble and Aptar Pharma, can assist untrained patients with at-home, self-administering of drug injections. The AdhereIT device accurately detects an injection event and lets the patient know (through audio, visual, or haptic feedback) whether the injection was done correctly.

Ingestible capsules can provide an automatic technique to record the ingestion of medications at a single-dose resolution, thereby replacing *direct observed therapy* by *wirelessly observed therapy*. The Proteus Discover system developed by Proteus Digital Health is an ingestible capsule technology to measure the effectiveness of medication treatment, which in turn helps to achieve desired clinical outcomes by tracking patient's health patterns. The Proteus Discover system consists of ingestible sensors, wearable sensor patch, mobile app for the patient, and portal for the provider [60]. Once the medication and ingestible sensors are ingested, they reach the stomach. The gastric acid in the stomach is used by the ingestible sensors to generate a tiny biogalvanic signal which is transmitted to the wearable patch. The biogalvanic signal is not pH dependent and lasts for roughly 7 minutes. The remaining contents of the ingestible sensor are inactive and eliminated through the feces. The wearable patch records the information about the ingested medications (i.e., date, time, drug type, and its amount) and physiologic data (i.e., step count, sleep quality, and circadian rhythm). A digital record is sent to the mobile app and to the provider through ProteusCloud after prior permission from the patient. This helps to confirm the ingestion of medications on a dose-by-dose basis and hence compliance to the treatment. The data also helps to quickly identify the root cause for nonadherence such as lack of understanding, undesired side effects, or improper dosing schedules. The clinician can then make decisions on whether to initiate, optimize, or eliminate the medications. Proteus Discover was FDA cleared in 2012 and has been tested in different patient populations (e.g., those with heart failure, hypertension, or tuberculosis) [104–106]. In patients with hypertension, the Proteus technology helped the community pharmacists to identify the root cause of persistent hypertension. Wireless monitoring showed that two-thirds of the patients had pharmaceutical resistance while the remaining one-third took inadequate medications. A study on the adherence to tuberculosis therapy found that the Proteus Digital Health system may be able to correctly identify ingestible sensors with high accuracy, low risk, and high patient acceptance [105].

A collaboration between Proteus Digital Health and the Japanese Pharmaceutical Company Otsuka led to Abilify's MyCite, which is the first digital medicine approved by the FDA to treat schizophrenia. The drug, Abilify (or Aripiprazole), is a prescriptive medication to treat mental disorders, such as depression, schizophrenia, bipolar disorder, and Tourette syndrome. MyCite is the smart pill version of the drug Abilify that embeds a tiny Ingestible Event Marker (IEM) sensor. This sensor is made of natural ingredients and passes naturally out of the body after use. Upon ingesting the smart pill, the IEM sensor transmits a Bluetooth signal to a wearable patch worn by the patient. The patch sends the information to a smartphone app and subsequently to a MyCite Dashboard which can be shared with a caregiver. The Abilify manufacturers have mentioned that FDA has evaluated and approved only the functions related to tracking the drug ingestion, and it is not known if Ability MyCite can help with patient compliance.

## 7. Ingestible Capsules for Gut Microbiome Analysis

Below are some examples of experimental ingestible capsules to collect gastrointestinal samples and track the microbiome within the gastrointestinal tract.

*7.1. Capsules for Collecting Samples of the Gastrointestinal Content.* The retrieval of physical samples from specific locations in the gastrointestinal tract can help us understand the composition and diversity of the microorganisms. The currently used stool test may not be reflective of the true intestinal health because fecal samples have shown to display greater bacterial richness and variation depending on factors such as stool consistency, medications, preexisting diseases, nutrition, and other personal traits [21, 107]. Tethered endoscopy to retrieve samples of the gastrointestinal tract is invasive, inconvenient, and incapable of reaching the distal regions of the small intestine. As an alternative, the gut sample can be collected by using magnetic fields to actuate the opening or closing of capsules. In recent past, magnetic fields have been used to create actuation forces and torque to control the movement of medical catheters, tissue biopsy sampling, targeted drug delivery, flexible endoscopes, and 3D endoscopy with scanning catheters [62]. As an example, a drug delivery module was demonstrated using electromagnetic actuation and active locomotion of the capsule through the gastrointestinal tract [63]. The drug was initially encapsulated by axially magnetizing two soft magnetic rings. An external magnetic field was deployed to move the capsule to the target lesion. Thereafter, a pulsating radial magnetic field was used to open up the capsule for drug release. A plastic hinge ensured that the module returned to its initial shape [63].

Building on the principles of magnetic actuation, there is a recent interest in magnetically actuated devices for sampling the gut microbiome. There are multiple factors to consider while designing such capsules, such as the opening and closing mechanisms, techniques for tight sealing, material and structural strength, and sample crosscontamination. There is a tradeoff between capsule dimensions and the collected volume, where smaller capsules can pass conveniently through the gastrointestinal tract but would collect less sample volume, and vice versa. It is estimated that the DNA and RNA concentrations in the samples should be at least 100-500 ng/$\mu$L for reliable detection during downstream analysis [108, 109].

A soft mobile robotic capsule was developed for noninvasive sampling of the liquid digesta [62]. Each of the two halves of the capsule contained a disk magnet (6 mm in diameter, 3 mm in thickness) and were held together by a flexure hinge made from an elastomer composite reinforced with carbon fiber. Silicon rubber was the material for the seal rim, while polydopamine was used as the biocompatible coating for capsule's inner surfaces. The capsule was normally closed when ingested from the mouth. An external magnetic field generated a torque to open up the two halves of the capsule to encapsulate the gut content. An opposite magnetic torque was generated to create an intermagnet force that sealed the capsule back to its normally closed configuration. With a capsule volume of 42 mm$^3$, the device collected 18 to 61 mg of gut content from pigs during *in vivo* studies. This amount of content was sufficient as 10 mg of material yields around 1.5 μg of DNA and 4 μg of RNA [110, 111]. Another study demonstrated the use of a highly absorbent hydrogel to first collect the gut microbiome and then automatically seal the capsule by mechanical actuation of the hydrated gel [112]. The capsule dimensions were 9 mm in diameter and 15 mm in length. A pH-dependent enteric coating was made with cellulose acetate phthalate (CAP) to keep the capsule closed at pH values less than 5.0 and to slowly degrade at higher pH. The inner surface of the capsule was treated with plasma and PEG to create a superhydrophilic surface for easy wicking and flow of gut content. The performance of the capsule in extracting and retaining viable bacteria was tested *in vitro* using *E. coli* solutions [112].

A soft capsule endoscope was constructed for the fine-needle biopsy of submucosal tissues in the stomach [113]. The capsule had an internal permanent magnet to respond to changes in the external magnetic field and indicate its physical location. A series of electromagnets generated the magnetic field to control the orientation and rolling motion of the capsule until the target location was reached. An endoscope camera gathered positional data and other information on the stomach, such as the detection of lesions. Upon reaching the destination, magnetic torque was applied to properly orient the capsule. Thereafter, magnetic force was applied to engage a thin, hollow needle to retrieve and store a tissue sample using the fine-needle aspiration biopsy technique. After sample collection, the needle was retracted with a tether attached to the capsule. *Ex vivo* experiments were conducted in a porcine stomach model with phantom tumors located underneath the outer stomach wall and produced 85% yield with the capsule robot [113].

*7.2. Capsules for Bacterial Biomarkers and Nucleic Acids.* Ingestible capsules have been housed with engineered microbes to detect health-related biomarkers and with electronic readout modules to wirelessly communicate the sensing signal at real-time [114]. This approach has benefits over current methods of probing the gut microbiome using plate analysis of urine and stool samples [12, 115, 116].

In one study, genetically engineered probiotic *E. coli* bacteria was leveraged to construct two biological sensors (thiosulfate sensor and tetrathionate sensor) which can detect colon inflammation and possibly lead to new methods to study gut microbiota pathways [117]. In another study, an ingestible microbioelectronic device (IMBED) capsule was developed to detect gastrointestinal bleeding events. The target biomarker was heme that was released from lysed red blood cells. The capsule combined engineered probiotic bacterial sensor with a bioluminescence detection circuit and wireless communication electronics [118]. The probiotic *E. coli* bacteria was housed within individual wells covered by a semipermeable membrane and was engineered to internalize the extracellular heme through an outer membrane transporter, ChuA. Upon detecting the target biomarkers, the heme-sensing genetic circuit was triggered to allow the bacterial luciferase operon, *luxCDABE*, to emit bioluminescence. The optical signal was read by phototransistors, converted to digital codes by a nanowatt-level luminometer chip, and transmitted wirelessly for data analysis. *In vitro* studies were conducted on a porcine model for gastrointestinal bleeding to correlate the dose-dependent heme input from whole blood to the luminescence production. An orogastric tube was used to deposit the IMBED capsule in pig's stomach, and photocurrent data was recorded for the next 2 hours. The above method was expanded to sense thiosulfate and acyl-homoserine lactone (AHL) that are biomarkers for gut inflammation and infection, respectively.

In a separate study, an ingestible biomolecular sensing system was developed for the detection of microbiome nucleic acids in the gastrointestinal tract [119]. The device fitted within the FDA-approved capsule size of 000. The capsule system was equipped with a fluorescence-based sensor, UV LED, optical waveguide, 1.55 V 12.5 mAh battery, and bidirectional communication features (124 μW wireless receiver and 915 MHz FSK/OOK transmitter). The CMOS fluorescent sensor chip had a 15-pixel array where each pixel had a 100 pM DNA detection limit. This detection limit was suitable for detecting the gut microbiome distribution *in vivo*.

## 8. Commercialized Platforms for Video Capsule Endoscopy

Video capsule endoscopy is the specific implementation of ingestible capsules designed to image the gastrointestinal tract [120, 121]. Compared to tube endoscopes which are invasively inserted through the oral or rectal cavity, capsule endoscopes have the capability to reach and clearly capture images inside the small bowel without causing discomfort to the patient [38, 122, 123]. Capsule endoscopes have been used to diagnose abnormal conditions in the gastrointestinal tract, such as chronic refractory constipation, celiac disease, and Crohn's disease [9]. The two common applications of small bowel capsule endoscopy (SBCE) are the diagnosis of suspected Crohn's disease and obscure gastrointestinal bleeding (OGIB) [124]. In fact, video capsule endoscopy is the single most effective endoscopic procedure to identify the source of OGIB [125]. Colon capsule endoscopy is commonly used for incomplete colonoscopy and patients unwilling to try regular colonoscopy [124]. Over the past two decades, video capsule endoscopy has become safer, and the rate of adverse events has declined as concluded from a survey of 402 published studies involving 108,000 capsule endoscopy procedures [126]. The capsules for endoscopy are generally equipped with a high-resolution camera, LED light source, wireless transmitter, different sensors, and batteries [127]. The patient may wear a portable receiver to acquire, store, and even display the images captured.

*8.1. Examples of Video Capsule Endoscopes and Rapid Reading Software.* Below are some examples of commercial platforms for video capsule endoscopy. Table 4 lists the companies for video capsule endoscopy, sample products, dimensions, features, specifications, and clinical applications.

TABLE 4: Commercialized products for video capsule endoscopy.

| Company | Sample products | Size ($D \times L$), weight | Features and technical specifications | Clinical applications |
|---|---|---|---|---|
| Medtronic | PillCam SB3 capsule, SB3 sensor belt, SB3 sensor array | 11 mm × 26 mm, 2.9 g | 156° FOV, 2-6 fps, 256 × 256 resolution, 4 LEDs, 3-lens system, 8 hrs battery life, automatic light exposure, high-quality imaging, adaptive frame rate, CMOS image sensor | Direct visualization of small bowel; monitor lesions from obscure bleeding, iron-deficiency anemia, and Crohn's disease |
| | PillCam Patency Capsule | 11 mm × 26 mm, 2.9 g | Dissolvable components, RFID tag detectable by X-ray, starts to dissolve 30 hrs after ingestion | Accessory to PillCam SB3 capsule; verify an unobstructed GI tract prior to capsule endoscopy |
| | PillCam UGI system | 11 mm × 26 mm, 2.9 g | 18-35 fps, variable frame rate technology, dynamic technology to pick clinically relevant information | Visualization of the upper GI tract; detect gross blood in esophagus, stomach, duodenum |
| | PillCam colon 2 capsule, colon 2 sensor belt, colon 2 sensor Array | 11.6 mm × 32.3 mm, 2.9 g | 336° FOV, 4 or 35 fps, 256 × 256 resolution, 10 hrs battery life, 2 camera heads, 4 white LEDs on each side, 434.1 MHz | Visualization of colon; detect colon polyps; alternative to colonoscopy |
| | PillCam Crohn's capsule, sensor belt, sensor array | 11.6 mm × 32.3 mm, 2.9 g | 336° FOV, 4 or 35 fps, 256 × 256 resolution, 10 hrs battery life, two heads, 4 white LEDs on each side, 434.1 MHz | Visualization of small bowel and colonic mucosa; detect subtle mucosal lesions |
| Intro-Medic | MiroCam capsule, MiroCam Navi, MiroCam green, MiroView 4.0 | 11 mm × 24 mm, 3.25 g | 150° FOV, 6 fps, 320 × 320 resolution, 11 hrs battery life, 6 white LEDs, magnetic controlled navigation, CMOS imager | Detection of bleeding, vascular lesions, ulcerations, polyps |
| Jinshan | OMOM capsule, OMOM Vue software, belt antenna | 13 mm × 28 mm, 6 g | 140° FOV, 2 fps, 0.1 mm resolution, 10 hrs battery life, CMOS image sensor, image enhancement technology | Low-cost capsule endoscopy; detect suspicious GI bleeding, angiodysplasia, tumors, ulcers |
| CapsoVision | CapsoCam Plus, CapsoView, CapsoCloud | 11 mm × 31 mm, 3.8 g | 360° FOV, full panoramic view, 20 fps, 15 hrs battery life | Direct lateral view of small bowel mucosa, lesions and other gut abnormalities |
| RF System Lab | Sayaka | 9 mm × 23 mm, N/A | 360° FOV, 30 fps, 410000-pixel CCD camera, 75 time magnification on-screen, mosaic-based moving image technology, battery-less wireless power transfer | View whole inner surface of GI tract; monitor peptic cell movement; locate secular distortion |
| Olympus Medical | EndoCapsule EC-10, small intestinal capsule, recorder set, antenna units | 11 mm × 26 mm, 3.3 g | 160° FOV, 2 fps, 0-20 mm depth of field, 12 hrs battery life, intelligent reading algorithm, advanced optics, CCD imager | Small intestine endoscopy; detailed observation of small intestine mucosa and abnormalities |
| Sonopill Programme | Sonopill robotic capsule endoscopy (RCE), MultiPill | 21 mm × 39 mm, N/A | Closed-loop magnetic control, ultrasound transducers, LED, camera, echo detection, accuracy of 2 mm and 3° | Robotic ultrasound endoscopy; scan GI tract; identify bowel features |
| AnX Robotica | NaviCam SB, stomach, UGI, colon capsule systems, ESView, engine | 11.8 mm × 27 mm, 5 g | Magnetic guidance, 360° automated scanning, controlled navigation, dual camera, AI-based intelligent reading | Magnetically controlled capsule endoscopy, direct and controlled visualization of GI disorders |
| Check-Cap | C-Scan Cap X-ray capsule, C-Scan Track, C-Scan View | N/A | Low dose X-ray beams, detectors for X-ray fluorescence photons and Compton back-scattered photons, integrated positioning and control, cloud-based analysis, 3D maps | Detect colorectal polyps, prevent colorectal cancer, alternative to colonoscopy |

(a) Medtronic offers a series of PillCam capsule endoscopy platforms for gastrointestinal care to detect abnormalities, monitor disease progressions, and evaluate treatment outcomes [124]. (i) The PillCam SB 3 System consists of the SB 3 capsule, SB 3 sensor belt, SB 3 sensor arrays, and SB 3 Recorder. The PillCam SB 3 capsule helps visualize the small bowel and monitor lesions which may result from obscure bleeding, Crohn's disease, or iron deficiency anemia (IDA). Its image capture rate is changeable from 2 to 6 frames per second (fps). The PillCam Sensor belt, sensor arrays, and PillCam Recorder 3 are worn by the patient. The PillCam Reader Software v9 provides high-quality imaging and smarter video compilation with a simple and accurate mapping of pill's progress through the small bowel and colon. These system features help to identify pathologies with greater diagnostic confidence [128]. (ii) The PillCam UGI capsule allows direct visualization of the upper gastrointestinal tract to detect gross blood and diagnose bleeding events with confidence. It employs a variable frame rate technology where a faster frame rate (i.e., 35 fps) is used in the initial 10 minutes, and a slower frame rate (i.e., 18 fps) is used in the last 80 minutes of the procedure. (iii) The PillCam Colon 2 System consists of the Colon 2 capsule, sensor belt, sensor arrays, and recorder. The PillCam Colon 2 capsule provides direct and accurate visualization of the colon to identify pathologies, such as colon polyps which can progress to colon cancer. The PillCam Colon 2 System is suitable for patients where colonoscopy or sedation is not possible (e.g., those with bleeding in the lower gastrointestinal region). (iv) The PillCam Crohn's capsule helps to directly visualize the mucosa of the small bowel and colon to detect subtle mucosal lesions for accurate disease status. This capsule has an extrawide field of view (336 degrees) with two camera heads (168 degrees each).

Apart from the video imaging capsules, the PillCam Patency Capsule can be given to verify the gastrointestinal tract is open and unobstructed prior to capsule endoscopy. The patency capsule is made of dissolvable components and a RFID tag that is detectable with X-rays. If the patency capsule leaves the body before dissolving (i.e., within 24 to 30 hours after ingestion), the gastrointestinal tract is confirmed passable or "patent" for the PillCam video capsules. Otherwise, if the patency capsule is retained for a longer period (i.e., greater than 30 hours), it starts to dissolve and disintegrate suggesting a possible obstruction in the gastrointestinal tract.

(b) EndoCapsule 10 System was developed by Olympus Medical. Its EC-S10 capsule has a 160° field of view and long observational time (i.e., up to 12 hours). The EndoCapsule 10 software was built upon the Olympus rich experience in optical-digital technologies to produce high-quality images of the small intestine with less noise and less halation. The software has a Red Color Detection function to detect active bleeds and ulcers, 3D track function to visualize capsule's progression through the small intestine, Omni Mode function to accelerate the reading time with safe detection, and Image Adjustment function to enhance tiny mucosal structures and alter color tones. An all-in-one recorder combines the receiver with a convenient belt-style antenna unit and the viewer to capture or play back endoscopic images. A recent multicenter study across nine European centers tested the efficacy of Omni Mode rapid reading software on patients with possible anemia or gastrointestinal bleeding [129]. A total of 2127 gastrointestinal lesions were examined using the normal mode (where every captured image was reviewed) or the Omni Mode, where 647 lesions were missed by normal mode and 539 lesions missed by Omni Mode. While the clinical accuracy were similar in both modes, the use of the Omni Mode reduced the mean reading time of endoscopic images by 40% compared to the normal mode (42.5 minutes) [129]. Another multicenter study involving 63 patients with obscure gastrointestinal bleeding compared the efficacy of EndoCapsule and PillCam SB capsules on the same patients [125]. The study found that both the capsules were safe and displayed comparable diagnostic yield with a subjective difference in the image quality in favor of the EndoCapsule

(c) MiroCam Capsule Endoscopy System was developed by IntroMedic as part of its Human Body Communication Technology (Figure 5). The MiroCam 1600 capsule has a higher frame rate (i.e., 6 fps) to provide a clear view of the small bowel while the MiroCam 2000 capsule is double tipped and with a lower frame rate (i.e., 3 fps) to provide more coverage of the bowel. The MiroCam Navi Controller uses magnetic forces to control the movement of capsules from the esophagus into the duodenum. The MiroCam Receiver is attached to a disposable belt to view and store the endoscopic images at real-time. The MiroView 4.0 software has a new Express View feature to remove redundant images and detect abnormal lesions, ulcers, bleeding, and polyps

(d) NaviCam Capsule System by AnX Robotica Corporation does not rely on natural peristalsis but is rather maneuvered by magnetically controlled robotics. The NaviCam capsule has a permanent magnet inside its dome and is activated by a capsule locator device that uses a magnetic sensor to ensure the NaviCam capsule is still present inside the patient. There is an ergonomic sensor belt and data recorder. The capsule is guided through the gastrointestinal tract by an external C-arm type magnet

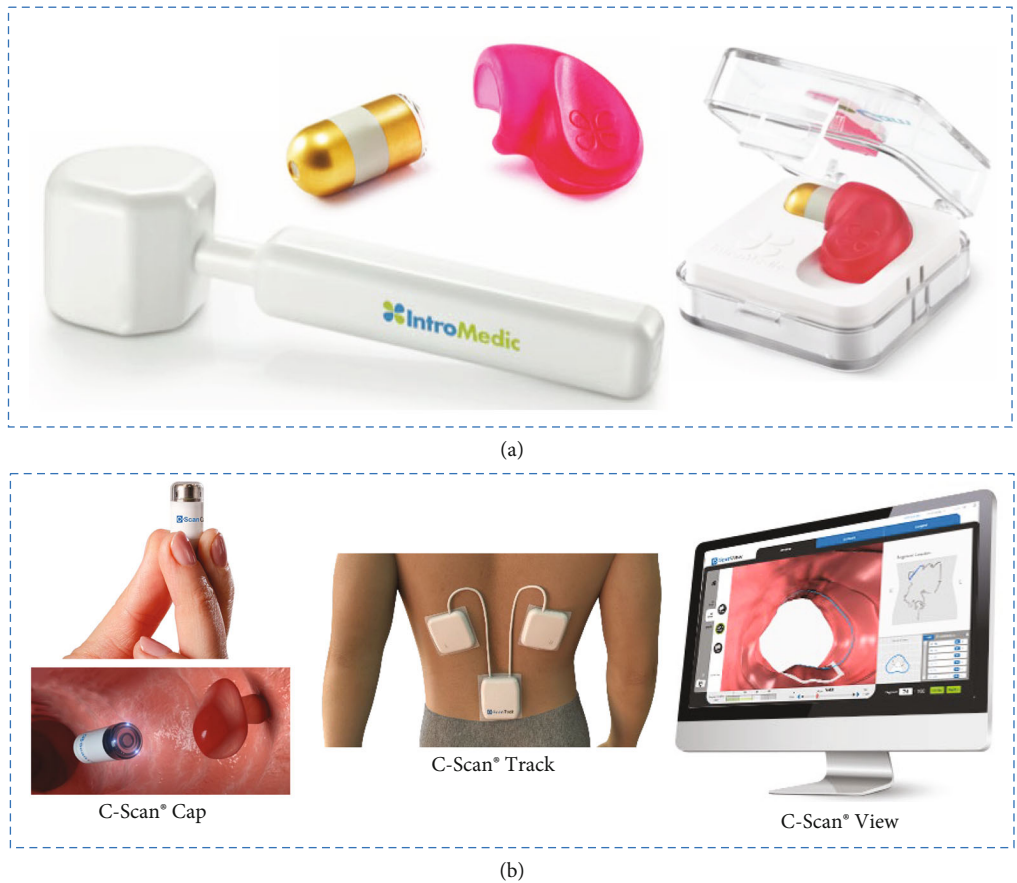

Figure 5: Video capsule endoscopy and X-ray imaging. (a) The MiroCam capsule endoscope system by IntroMedic is designed to take images of the small bowel mucosa to detect abnormalities. (b) The C-Scan Cap by Check Cap scans the gastrointestinal tract using ultralow-dose X-ray beams. The C-Scan Track consists of three patches worn on user's back for control, monitoring, and recording of the capsule information. The C-Scan View is their cloud-based analysis suite for the 3D reconstruction of the colon and detection of colorectal polyps. The images were reproduced with permission from IntroMedic (a) and Check Cap (b).

robot having 5 degrees of freedom (i.e., 2 rotational and 3 translational) and an adjustable magnetic strength up to 200 mT. Two separate joysticks are used by the operator to control the motion of the capsule in the 2 rotational axes (i.e., vertical and horizontal directions) and 3 translational axes (i.e., up/down, left/right, or forward/backward). An automatic reset button brings the capsule to its default position. The NaviCam System is adapted to image the gastric mucosa, duodenum, small bowel, and colon with 360° automatic scanning capability and real-time view (e.g., NaviCam Stomach Capsule, NaviCam SB, NaviCam UGI, and NaviCam Colon). The ESView software with ProScan Reading Support and NaviCam Engine employs artificial intelligence algorithms to screen out abnormal lesions with high sensitivity

(e) CapsoCam Plus capsule endoscopy system by CapsoVision provides 360° panoramic view of the small bowel to directly visualize the mucosa to identify any lesions or abnormalities (Figure 6). There is no need for belts or wires, and all data is stored within the on-board flash memory system. The CapsoRetrieve Magnetic Retrieval Wand is used to easily remove the capsule after its exit from the gastrointestinal tract. The CapsoView 3.6 diagnostic software provides various playback controls to view examination videos, a Red Detection system to identify suspected bleeding areas, suggested landmarks, Advanced Color Enhancement (ACE), duo and panoramic view modes, customizable report generation, and a reference library of related pathologies/landmarks to help physicians in the diagnosis. The CapsoCloud is a cloud-based data management software (hosted by Amazon Web Services) that gives the ability to stream and download CapsoCam Plus videos on a portable iOS or Android devices

(f) The OMOM capsule endoscope by Jinshan Science & Technology has an ergonomic design with a light and compact enclosure. The recorder has a belt antenna system to pair up with the capsule and a display to view the endoscopic images at real time. The

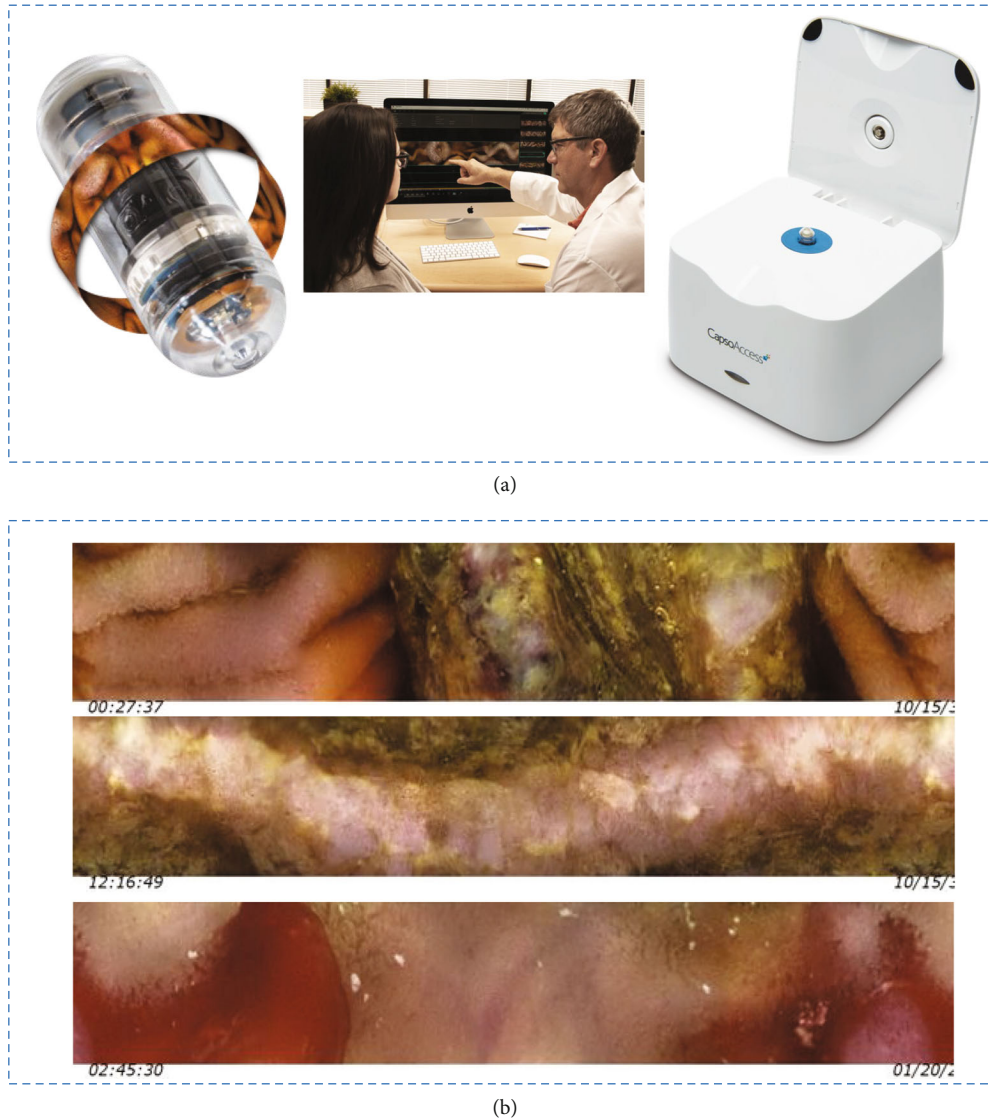

Figure 6: Rapid reading software by CapsoVision. (a) The CapsoCam Plus System uses four lateral cameras to capture full 360° panoramic view of the small bowel mucosa. The high-resolution images are reviewed on the CapsoView 3.6 diagnostic interface that has several features, such as multiple viewing modes, playback adjustments, easy archiving, Red Detection system, suggested landmarks, and report generation. (b) Examples of image streams of the small bowel mucosa captured by the CapsoCam Plus and reviewed by the CapsoView 3.6 diagnostic interface. The images were reproduced with permission from CapsoVision.

OMOM Vue Smart software provides a simple and easy workflow experience for the user to control video play options, skim through images, identify similar traits, and compare multiple cases in parallel. Image color enhancement (ICE) is provided by the RGB mode to clearly view the capillaries, mucosa, and areas of suspected bleeding. The OMOM platform is cost-effective compared to its peers, which makes it suitable for large-scale clinical studies on diagnostic efficiency [130]

(g) Norika and Sayaka endoscopic capsules were developed by RF System Lab. While the older Norika capsule has the camera fixed at one end, the newer Sayaka capsule puts its camera on the side and added a tiny stepper motor to rotate the camera. The Sayaka capsule provides a 360° panoramic view and captures around 870,000 images during a typical 8-hour journey through the gastrointestinal tract. Both capsules are battery-free and powered wirelessly from a vest that sends out microwave signals. The raw images from different angles are transmitted to a nearby receiver and stitched together to recreate a rectangular gastrointestinal map using an image mosaicking technology

*8.2. X-Ray Imaging and Microultrasound Capsules.* Besides white light optics for video capsule endoscopy, X-rays and microultrasound have been used for imaging clinically relevant landmarks in the gastrointestinal tract as described below.

(a) The C-Scan Cap by Check-Scan is an ingestible X-ray imaging device designed to screen for precancerous polyps and colorectal cancers (Figure 5) [131]. This noninvasive test serves as an alternative to standard colonoscopy and does not require bowel preparation prior to the procedure (which is a key deterrent for colorectal cancer screening). Once the C-Scan Cap capsule is swallowed and reaches the colon, it generates low dose of X-ray beams (mean radiation dose of 0.04 mSv) which interact with the ingested contrast agents, tissue walls, and contents of the colon. The X-ray detectors within the capsule constantly measure the number of reflected X-ray fluorescence photons and the Compton backscattered photons to estimate the distance between the capsule and inner wall of the colon. The raw images from the C-Scan Cap are transmitted and stored onto the C-Scan Track comprising three patches worn on patient's back. After the procedure, the C-Scan Cap capsule is excreted naturally. Thereafter, the C-Scan View software analyzes the scanned data and create 3D maps of colon's inner surface to detect precancerous polyps. A feasibility study tested the safety and efficacy of the C-Scan Cap for the detection of colonic polypoid lesions and masses [132]. The capsule results from 45 patients were compared with fecal immunochemical test (FIT) and ensuing colonoscopy. The data indicated a 78% capsule sensitivity (when >50% of the colon surface area was imaged) and 90% capsule specificity

(b) The Sonopill capsule combined the magnetically controlled robotic capsule technology with microultrasound imaging for the diagnosis of gastrointestinal cancers. Traditionally, the gold standard for colorectal cancer screening is standard flexible endoscopy, which presents a higher procedural complexity, severe discomfort to patients, and inability to access all areas of the colorectal region [133]. The Sonopill program was started from a 5-year, $10 M grant to four U.K university partners (i.e., Glasgow, Dundee, Heriot-Watt, and Leeds). Its capsule consisted of a CMOS camera, electronic circuitry, two microultrasound transducers, and a permanent magnet (neodymium iron boron (NdFeB)) [7]. The system used microultrasound feedback control to precisely maneuver the robotic capsule and scan the surface to capture ultrasound images. The microultrasound closed loop feedback was used to inform about the robot motion, and thereafter, robot's motion was used to reposition the microultrasound transducer. An autonomous echo-finding algorithm was implemented to regularly adjust capsule's pose and its acoustic coupling with the tissue. The output data was combined to reconstruct the substrate and detect features with a positional accuracy up to 1-2 mm in living porcine models

*8.3. Imaging-Guided Therapeutics and Gastrointestinal Surgical Procedures.* Apart from visualizing the gastrointestinal tract for clinical diagnosis, capsule endoscopy systems have been employed to deliver drugs at specific locations of the gastrointestinal tract, assess tablet disintegration *in vivo*, perform fine-needle aspiration biopsy, and treat gastrointestinal hemorrhaging. In a study for controllable drug delivery, a magnetic drug delivery capsule was developed consisting of a drug chamber, magnets, copper coils, and electronic modules [134]. When the copper coils were electrically activated, the magnetic force detached the drug chamber from capsule's body to release the drug in its vicinity. In a separate study to determine the timeline of tablet disintegration within the gastrointestinal tract, paracetamol tablets were retrofitted to an ingestible capsule and administered to beagle dogs [135]. The time taken for tablet's *in vivo* dissolution was obtained from its plasma concentration profile and visual confirmation from the camera capsule. This technique provides a comfortable alternative to balloon endoscopy.

Gastrointestinal hemorrhaging is an umbrella term for any form of bleeding that originates in the gastrointestinal tract. The symptoms depend on both location and severity but may include bright red blood in vomit, cramps in the abdominal region, and shortness of breath. In the United States, over 100,000 people suffer from upper gastrointestinal bleeding, and men are twice as likely to have upper gastrointestinal bleeding [136]. One study showed the use of ingestible capsules to treat gastrointestinal hemorrhaging through balloon tamponade effect [137]. The capsule was equipped with a silicone balloon that inflated by the gaseous reaction created by mixing an acid (acetic acid, 5% concentration) and a base (sodium bicarbonate). The gaseous byproduct (carbon dioxide) was funneled to the balloon to inflate it against the gastrointestinal walls, thereby creating pressure and blocking the source of hemorrhage. The capsule consisted of three different sections linked by flexible joints. *In vivo* experiments on a sedated pig showed that the balloon was immediately inflated upon reaching the hemorrhage. Another capsule technique for treating gastrointestinal hemorrhage involved the wireless deployment of surgical clips in order to close the wound [138]. The capsule had a cylindrical shape (diameter = 12.8 mm and length = 33.5 mm). It was equipped with 4 magnets on the external surface for magnetic steering, a nitinol (nickel-titanium alloy) clip to close the wound, an electromagnetic motor, and a communication module [138]. *In vivo* experiments on pigs showed the successful blockage of internal hemorrhage.

## 9. Materials and Energy Sources for Next-Generation Ingestible Capsules

Next-generation ingestible capsules are being developed using biocompatible and biodegradable materials, along with biocompatible batteries and alternative energy sources as discussed below.

*9.1. Biocompatible Materials, Edible Food, and Nutrients.* Biocompatibility describes the ability of the capsule system to exist in harmony with the gastrointestinal system. By fabricating biocompatible capsule components, the risks of prolonged capsule retention can be mitigated [122, 139]. The physical dimensions and mechanical characteristics of the capsule also dictate its biocompatibility with the gastrointestinal tract. For instance, the maximum size of capsule's rigid outer body is limited to the diameter of the smallest passage within the gastrointestinal tract (i.e., 12.8 ± 7 mm). This spatial constraint sets an upper limit on capsule's functionality. As such, flexible biocompatible materials are sought that can simultaneously address the issues of size limitation and prolonged capsule retention.

Research into biocompatible electronics for capsules has garnered considerable interest [14, 140–143]. A team of researchers printed active electronic devices on edible substrates using temporary tattoo paper, biocompatible conducting polymers, and inkjet circuit printing technology [144]. The p-type and n-type organic field-effect transistor-based circuits (OFETs) were fabricated and transferred onto the surface of a hard gelatin capsule by soaking in water and pressing them onto the target surface. The submicron thick ethyl cellulose film within the temporary tattoo paper served as the transferrable substrate and as the gate dielectric of the OFETs. The biocompatible materials used to form the transistors were poly(3-hexylthiophene) (P3HT) and different donor–acceptor copolymers. While this work realized a simple logic inverter, it showed the possibility of using printable edible electronics for ingestible capsules.

A commercial example of an ingestible capsule system that incorporates biocompatible materials is the Ingestible Sensor System (ISS) developed by Proteus Digital Health to test the patient adherence to oral medications. Its Ingestible Event Marker (IEM) sensor has an integrated circuit housed within an edible, cellulose disc. Thin layers of copper and magnesium (10 $\mu$m) form the biogalvanic battery on the surface of the IEM sensor. The extractable amount of copper and magnesium from a single IEM sensor was minimal compared to the daily allowable intake levels (0.3% and 0.003%, respectively) [145]. Ethyl citrate was extracted from the IEM sensor as the organic species resulting from the breakdown of triethyl citrate (considered a GRAS food additive and inert material).

The design of ingestible capsules has benefitted from the emerging trends in transient electronics where the material and devices can physically degrade in a controllable manner. Within the gastrointestinal tract, the degradability and depolymerization traits of the transient electronics can be tuned based on its transit time or external triggers (e.g., pH, temperature, light, or solubility in solvents found in the gastrointestinal tract). In one study, transient antennas were developed in a degradable composite film made of water-soluble poly (vinyl alcohol) (PVA) [146]. The dissolution rate and dielectric properties of the PVA film was tuned by doping with $TiO_2$ nanoparticles. The fabricated patch antennas degraded within one hour of water immersion and were space-efficient to fit within capsules. Besides antennas, flexible substrate films have been chosen that conform to the capsule walls. Biodegradable films (e.g., polyanhydrides) have been tested as substrates that can partially or completely degrade at finite time points. The biocompatibility of substrates reduces the health and safety risks associated with long-term capsule retention beyond 24 to 48 hours. Flexible and stretchable components, such as antennae, can be packaged in miniaturized form factors to facilitate the initial swallowing and transit and expanded to larger form factors upon reaching the small bowel or colon [8, 147]. On the other hand, sensors and microelectronics are often compressed into mechanically rigid structures to save the real estate within capsules.

The materials for transient electronics are being chosen based on their inherent characteristics, such as permeability, degradation rates, morphology, hydrophobicity or hydrophilicity, chemical properties, and mechanical properties of the polymer [115, 148, 149]. In a study on bioresorbable silicon electronic sensors, devices were fabricated on silicon nanomembranes, thin $SiO_2$ layers, and magnesium foils and molybdenum interconnects that can dissolve or disintegrate in a time-controlled manner [140]. Another study demonstrated a range of functional transient devices fabricated on polyanhydride substrates (~124 $\mu$m thick) where transience was triggered by moisture and time-controlled by the polymeric chemistry [150]. The anhydride groups in the polymeric substrate absorbed the surrounding moisture to cause its hydrolysis and eventually degrade into a viscous liquid. Various transient electronic devices were fabricated from metals (e.g., copper, nickel, and aluminum), semiconductors (e.g., indium gallium zinc oxide (IGZO)), and dielectrics (e.g., magnesium oxide) [150].

The material toolkit for next-generation ingestible capsules can be constructed from a number of constituents in edible food and nutrients (such as minerals, vitamins, inorganics, organics, and food additives). Table 2 lists the different categories of biocompatible materials for next-generation capsules: substrates, fillers, coatings, conductors, electrodes, and electrolytes. To help choose the appropriate biocompatible material, a system of nutrition recommendations and dietary reference intake (DRI) can be used as reference, which is maintained by the U.S. National Academies and Institute of Medicine. As a note, the recommended dietary allowance (RDA) indicates the sufficient daily intake level of essential nutrients, while the tolerable upper intake level (UL) refers to the highest intake level of nutrients considered safe with no side effects. Materials tabulated by the US RDA include minerals, organic molecules, polymers, and inorganic materials. Minerals that have tabulated RDA and electrical properties can be used to create essential components within ingestible electronic devices, such as the anode, cathode, electrode, flow cell ions, and active layer. The body can tolerate a threshold of most materials before the concentration within the blood stream and tissue becomes detrimental [151]. In addition, FDA has classified certain chemicals and substances as "Generally Recognized As Safe" (GRAS) for common human consumption. The preapproved safety of GRAS materials makes them good candidates for use in ingestible capsules as they have a lesser path of resistance for clinical trials.

*9.2. Batteries and Alternative Energy Sources.* Silver-oxide button batteries are popular energy sources for portable electronics because of their high energy-to-weight ratio. However, silver-oxide button batteries occupy significant real estate within ingestible capsules and often become the deciding factor for the capsule size. The shape and size of these batteries pose potential obstruction and health risks in children when swallowed [152, 153]. The alkaline solution within ingested silver-oxide button batteries can cause severe tissue damage in the mouth, vocal cord, trachea, or esophagus [154, 155]. Lithium cells are more dangerous than silver-oxide batteries, especially when ingested by children [153, 155]. When the alkaline electrolyte of the lithium cell leaks into the stomach and comes in contact with the electrolyte-rich stomach fluid, undesired electrical current is generated. This results in the local production of hydroxide ions and pH change within the stomach, leading to necrosis and potential fatalities in children [152, 156, 157].

Solid-state cells and biocompatible batteries are potentially safer than conventional button batteries because there is no leakage of toxic electrolytes or associated risks of internal tissue damage. Biocompatible batteries have a variety of material choices for the anode (e.g., magnesium, zinc, magnesium-aluminum-zinc alloy, or melanin), cathode (copper, cupric chloride, iron, platinum, or manganese dioxide), and electrolyte (e.g., sodium chloride, phosphate buffer, saliva, or simulated gastric fluid). As an example, a biocompatible galvanic cell was demonstrated where gastric acid and intestinal fluids served as the electrolyte [158]. A redox reaction occurred between the dissolving metallic anode (zinc or magnesium) and an inert cathode (pure copper metal). *In vitro* experiments indicated that the zinc anode was better suited for longer term use because the magnesium anode was prone to greater corrosion and shorter lifetime (<24 hours). Their energy harvesting system was tested in a porcine stomach to power the temperature-sensing electronics housed within the capsule and wirelessly transmit the signal to a base station (as a 920 MHz, FSK modulated signal). The average amount of zinc absorbed in the body from this galvanic cell (i.e., 27 mg/day) was estimated to be below the recommended upper limit (i.e., 40 mg/day). A drawback of biocompatible batteries is that the battery performance can fluctuate due to variations in the environmental conditions of the gastrointestinal tract.

Alternative energy sources have been explored for ingestible capsules, including biofuel cells, nuclear batteries, optical charging, and transducers of various types (e.g., thermoelectric, piezoelectric, electrostatic, electromagnetic, and ultrasound). Within the field of chargeable ingestible batteries, a polysaccharide gel battery based on the Gibbs-Donnan effect (i.e., Donnan dialysis) was recently demonstrated to fight harmful bacteria in the oral cavity [159]. In their work, electrochemical potential was generated by ionic diffusion from potassium chloride saline gradients within agarose gels. The battery and triboelectric nanogenerator were created from biocompatible materials, such as agarose gel, platinum, polyethylene terephthalate (PET), polytetrafluoroethylene (PTFE), chitosan/glycerol film, and silicon. An optimized battery configuration produced an open circuit potential up to 177 mV and the nanogenerator was capable of supplying 300 mV. A system using two such batteries in series was demonstrated as an antibacterial electrostimulation device for the oral cavity. Their test results showed that 90% of *E. coli* were inhibited after 30 minutes of treatment.

Besides on-board energy sources, remote powering of ingestible capsules using wireless power transmission have been investigated by researchers [59, 160]. In one approach, a Helmholtz transmitting coil was attached to the patient and a ferrite-core three-dimensional receiving coil was placed in the capsule. Electromagnetic fields were used to energize the capsule, providing sufficient power (i.e., 330 mW to 500 mW) to the capsule for most of its functions. The design and size of the antenna determined the frequency of optimal transmission. As a drawback, wireless power transmission is critically dependent on the distance between the source and receiving antenna and capsule's relative orientation to the incoming radiation. There is significant signal loss by the time the electromagnetic waves reach the deep gastrointestinal tissues (where the ingestible capsules typically operate).

## 10. Discussion: Present Challenges and Outlook

First and foremost, it is challenging for a clinic or hospital to constantly assess and improve the endoscopic service performance measures with root cause analyses of major events and feedback to their gastroenterologists and medical staff [40]. Moreover, these procedures are not appropriate for patients having gastrointestinal obstructions, peptic stricture or ulcer disease, urethral strictures, urinary fistulas, uncontrolled diabetes, swallowing disorders, prior gastrointestinal surgery, and implanted electromedical devices. To avoid electromagnetic interference during the time from capsule ingestion to excretion, it is advised to stay away from sources of powerful electromagnetic fields (e.g., MRI instrument). There may be frequent signal loss, difficulties in knowing the exact location of the capsule, or loss of battery power before excretion [76]. On the technical side, there are high costs associated with advanced microchips, software upgrades, and clinical validation that is expected to limit the market growth of smart pills [5]. Furthermore, it is difficult to reconstruct and democratize the sensing and recording circuitry from off-the-shelf components, which limits the number of commercial ingestible capsule products [161].

The procedures involving ingestible capsules have health risks, such as capsule retention, aspiration, obstruction, mucosal injury, internal bleeding, and irritation. In particular, capsule retention in the gastrointestinal tract can have significant health complications depending upon the preexisting conditions of the patient and nature of capsule materials [151]. A study found that capsule retention following video capsule endoscopy was predominantly found in individuals with bowel disease (1.4% retention rate) and Crohn's disease (13% retention rate) where the retention of capsules often resulted in gastrointestinal obstruction [162]. A survey of 22,840 procedures using small bowel capsule endoscopy showed a significant retention rate for common indications such as obscure gastrointestinal bleeding (1.2%), suspected Crohn's disease (2.6%), and neoplastic lesions (2.1%) [163]. Other underlying conditions for capsule retention are obstructive tumors and diaphragm disease from the side effects of anti-inflammatory drugs [164]. Extended retention of a rigid-shell capsule can result in lumen perforation and capsule disintegration with undesired toxicity from released materials. In such cases of prolonged capsule retention (especially having small-bowel strictures), the retrieval of capsule is assisted by conventional deep enteroscopy, push enteroscopy, or motorized spiral enteroscopy [165]. To address the risks of capsule retention, Medtronic has devised medical procedures to evaluate the safety of patients before proceeding with video capsule endoscopy. For instance, its PillCam Patency Capsule is ingested by the patient prior to video capsule endoscopy to check for known obstructions and is possibly excreted within 30 hours. A study of 38 patients with risk factors of capsule retention showed that the PillCam Patency Capsule was naturally excreted in most cases, and video capsule endoscopy was safely performed afterwards in those patients [166].

Clinical testing of capsule prototypes on humans and laboratory animals is nontrivial, time-consuming, and expensive. Because of interpatient variability in terms of disease nature and treatment history, the patients who volunteer for clinical trials may not be representative of a target diseased group or the general healthy population [11]. It is difficult to correlate gastrointestinal parameters such as gastric emptying time, pH, gastrointestinal motility, colon contractions, and transit times both in healthy individuals and patients with preexisting conditions (e.g., obesity, chronic constipation, and diabetic gastroparesis) [79]. As an example, a validation study showed that temperature-sensing ingestible capsules have rapid temperature oscillations, especially in obese individuals, with temperature drifts compared to electronic and mercury thermometers [68]. The same study also revealed the current dearth of knowledge in relating the core body temperature to the body mass index, gender, age, and daily movement and activity and presence of food or anatomical location in the gut [68]. It is challenging to optimize the experimental parameters for different patients during video endoscopy, physiological sensing, or stimulation. In fact, any capsule system and its procedure have a large number of parameters with a range of possible variations, which makes it difficult to make informed guesses and predict outcomes, especially in patients with gastrointestinal disorders [11]. Synthetic gut models fail to emulate the structure and functions of live gastrointestinal tracts, such as peristalsis, bowel movement, and slippery mucosa layer. Peristalsis, for example, can force the capsules to go through certain sections of the gastrointestinal tract at different speeds, which may not allow sufficient time to take clear images. This can lead to lesions and other diagnostic pointers being overlooked or misread [167]. For video capsule endoscopy, this can prevent the reliable acquisition of high-resolution gastrointestinal images, reimaging and pinpointing the origin of a biological marker, and rapid reading of the acquired images with acceptable speed and accuracy [168].

The outlook for ingestible capsules involves addressing the present challenges and testing strategies for technical advancements and patient adoption. The field is moving towards a single, multifunctional, integrated circuit chip with in-built CMOS image sensors, physiochemical sensors, ASICs, memory units, transceivers, and antenna-in-package. Novel materials are being investigated with properties to encapsulate electronic components, perform energy harvesting, provide mechanical strength for faster transit, and have chemical properties for bio-absorption and degradability [151]. In the near future, closed-loop feedback will become essential to monitor the target parameters for gastrointestinal disorders at quasi real time and to automatically adjust the external control in a timely manner. The applications of small bowel capsule endoscopy will be expanded and validated for therapeutic functions such as for the treatment of obscure gastrointestinal bleeding or the detection of mucosal damage from drugs and NSAIDS injury [124]. As we better understand the intricacies of the gut-brain axis connection, the probiotics industry is emerging to be a big business for the assessment, exploitation, and rebuilding of the gut microbiota and commensal balance [107]. In this context, ingestible capsules could be developed to detect inflammatory proteins, hormones, or metabolites which are indicative viral infections or a modulated gut microbiome [169]. With expanding knowledge of sacral and vagal nerve pathways and underlying mechanisms, the emerging technologies of bioelectric neuromodulation and gastric electrical stimulation (GES) could be tailored and optimized for the treatment of gut inflammation, obesity, gastroparesis, and fecal incontinence [11]. The rising rates of gastrointestinal diseases and cancers suggest that we will see further subspecializations in therapeutic and diagnostic endoscopy with emphasis on procedural training, computing skills, and service quality [10]. The performance measures for endoscopic procedures will evolve with time and affect the organization, leadership, staffing, quality and safety, and patient involvement [40]. Artificial intelligence and personalized treatments will play a greater role in endoscopic therapeutics and diagnosis. Lastly, video capsule endoscopy could effectively complement existing procedures in surgery and radiology by establishing safe and high-quality out-patient procedures, reduced medical complications, and quicker diagnostics at lower costs [10].

## Data Availability

All the data is presented in the paper. Please contact S.P. for additional information.

## Conflicts of Interest

The authors declare that they have no competing interests.

## Authors' Contributions

S. P. planned the content and organized the structure of the paper. S. P., D. M., and L. B. communicated with the capsule manufacturers. All authors contributed to the figures, data tabulation, and discussion. S. P., D.M., and L. B wrote the first draft of the manuscript. All authors, including S. P., K. Y., and A. J., revised the subsequent drafts. Dylan Miley and Leonardo Bertoncello Machado contributed equally to this work.

## Acknowledgments

We are grateful to the capsule manufacturers for our discussions and giving permission to reproduce their technology figures here. This work was partially supported by the United States National Science Foundation (NSF IDBR-1556370) and U.S. Department of Agriculture (2020-67021-31964) to S.P.